\documentclass[preprint,12pt]{elsarticle}
\usepackage{graphicx}
\usepackage{bm}\usepackage{subfigure}
\usepackage{amssymb}
\usepackage{amsmath}

\journal{Physics Letters B}
\begin{document}
\begin{frontmatter}
\title{Isotropisation of Quadratic Gravity: Scalar and Tensor Components}

\author[a]{Daniel M\"uller}
\ead{muller@fis.unb.br}
 \author[b]{M\'arcio E. S. Alves}
 \ead{alvesmes@unifei.edu.br}
 \author[c]{Jos\'e C. N. de Araujo}
 \ead{jcarlos.dearaujo@inpe.br}
\address[a]{Instituto de F\'\i sica UnB,
 Campus Universit\'ario Darcy Ribeiro,\\
 Cxp 04455, 70919-970, Brasilia DF 
 Brazil}
 \address[b]{Instituto de Ci\^encias Exatas,
Universidade Federal de Itajub\'a, 37500 \\
Itajub\'a, MG,
Brazil }
\address[c]{Instituto Nacional de Pesquisas Espaciais - 
Divis\~ao de Astrof\'isica, \\Av. dos Astronautas 1758,
S\~ao Jos\'e dos Campos, 12227-010 SP,
Brazil}

%\toccontinuoustrue
\begin{abstract}
It is believed that soon after the Planck era, spacetime should have a semi-classical nature. Therefore, it is unavoidable to modify the theory of General Relativity or look for alternative theories of gravitation. An interesting possibility found in the literature considers two geometric counter-terms to regularize the divergences of the effective action. These counter-terms are responsible for a higher order derivative metric theory of gravitation. In the present letter we investigate how isotropisation occurs. For this reason a single solution is chosen throughout this article. We obtain perturbatively, by two different methods, that the tensor and scalar components emerge naturally during the isotropisation process. In this sense our result provides a numerical example to Stelle's well known result on classical gravity with higher derivates. Our entire analysis is restricted to the particular Bianchi type $I$ case.
\end{abstract}
\begin{keyword}
quadratic gravity\sep isotropisation\sep gravitational waves
\end{keyword}
%\PACS{98.80.Cq, 98.80.Jk, 05.45.-a}

\end{frontmatter}

\newcommand{\imsize}{0.85\columnwidth}
\newcommand{\halfsize}{0.40\columnwidth}

\section{Introduction}

A semi-classical theory considers the back-reaction of quantum
fields in a classical geometric background. This approach was proposed
a long time ago by de Witt \cite{DeWitt}, and is still a topic
of research (see, e.g., Ref. \cite{Hu}).

Differently from the usual Einstein-Hilbert action, in a semi classical theory,
the gravitational action results into a set of non linear partial differential equations of forth order, \cite{DeWitt}, see also \cite{liv}. 

It is well known that the linearized limit of these quadratic curvature actions involve a massless spin-2 field, a massive spin-2 ghost and a massive spin-0 field \cite{Chiba}. As pointed out by Chiba, this result can be obtained in three different ways, through the propagator \cite{Nunez}, direct analysis of the linearized fields \cite{Stelle1978} and through the no go theorem excluding consistent self couplings between spin-2 fields \cite{Deser}. 

In the quantized version of the theory, the presence of the ghost is unacceptable because it would either render the vacuum unstable, or destroy the unitarity of the theory \cite{Stelle}. 

On the other hand, in the context of the pure classical theory, which we are more concerned in this present work, the presence of the ghost is not so drastic as we explain in the following. For vacuum solutions, $T_{ab}=0$. Then, in the linearized, first order limit of the theory, the spin-2 ghost, spin-2 massless and spin-0 massive fields are all decoupled from each other and themselves. In other words, the linearized spin-2 ghost and all the other fields are entirely free fields in the absence of sources. So, in the classical sense, the coupling between the linearized fields will necessarily become important at higher levels of perturbation theory and depending on the classical source $T_{ab}$. In this sense, if the perturbation increases the interaction with the ghost spin-2 field introduces a mechanism of instability which is very well known. If the perturbation is sufficiently small the interaction is irrelevant and the linearized approach maintains itself. 

This higher order theory was previously studied by Starobinsky
\cite{S}, and more recently by, for example, Shapiro, Pelinson and others
\cite{Shapiro}, who focused their work on the homogeneous and
isotropic spacetimes. We refer the reader to a paper by Schmidt \cite{hjs}
for a review on higher order gravity theories in connection to cosmology.

Concerning studies of anisotropic spacetimes, Tomita \cite{berkin} was the first to investigate general
Bianchi type $I$ spacetimes. He found, for example, that the presence of an anisotropy contributes to the formation
of a singularity. Barrow and Hervik \cite{barrow-hervik}, in addition to addressing the anisotropic cases of
Bianchi type $I$, did the same for Bianchi type $II$. Their most interesting results are the derivations of exact
solutions for quadratic theories. Also interesting is the article by Saridakis in which the anisotropic Kantowski-Sachs cases are addressed for quadratic theories \cite{saridakis}.

For general anisotropic Bianchi type $I$ homogeneous spacetimes, the higher order
theory reduces to a system of nonlinear ordinary differential
equations with fourth order time derivatives, whose numerical solutions were obtained
in a previous paper \cite{sandro}. The stability of the Minkowski
spacetime, in this context, was also addressed and these authors obtained that there are
many initial conditions that result asymptotically in Minkowski space
\cite{daniel_sandro}. It is shown that there is a basin of attraction to the
Minkowski spacetime. Also, isotropisation can occur in de Sitter spacetime with
a non-zero cosmological constant. A more general, non diagonal line element in the Bianhi $I$ and Bianchi $VII_A$ cases was addressed, \cite{sandro}. The decomposition $3+1$ is given in \cite{Friedmann11}, with a specialization to the Bianchi $I$ case.

It should be mentioned that the isotropisation process for non-zero cosmological constant
is not a peculiarity of these higher order theories, since it also occurs in General Relativity, as proved
by Wald \cite{wald}. He showed that for $\Lambda>0$ all Bianchi models, except
for a highly positively curved Bianchi IX, the spacetime asymptotically becomes a de
Sitter spacetime.

Concerning the isotropisation process, a remarkable difference between
General Relativity and the higher order theories is that this last one
depends on the initial conditions.

In the present work our aim is to investigate how the isotropisation occurs and if
gravitational waves are generated in this process. For this reason an initial
condition is chosen in such a way that asymptotically spacetime becomes a flat spacetime.

%In an article by Matarrese et al \cite{matarrese}, it is argued that gravitational waves do not occur when the magnetic part of the Weyl tensor $H_{ab}=0$  vanishes, as in Bianchi $I$ geometries. It is worth noting
%however that these authors are not concerned with the distinction between gravitational waves and gravitational
%radiation in the Isaacson's sense \cite{Isaacson1}. As is well known, all solutions of Petrov type N, have a non vanishing $H_{ab}\neq 0$ tensor. Also, the space times with vanishing $H_{ab}=0$, vanishing shear, vorticity and pressure were named silent universes, see for example \cite{dsc}.

%It is worth recalling that there are many definitions of gravitational waves, see for example \cite{zakharov}.
%In a broader sense, and also non rigorous sense, many solutions can be understood as gravitational waves. For instance, the exact Bianchi $VII_h$ Petrov type N solution \cite{ogh} can be interpreted as a homogeneous gravitational wave. There is not a localized spatial source being the origin and therefore there is not a spatially compact wave front involving the source.

This paper is organized as follows. In section \ref{The field equations} we present the field equations and the spacetime metric considered in this article. We also reproduce the linearized fields found by \cite{Stelle1978} for completeness. In section \ref{Weyl} we obtain the Weyl scalars from the full numerical solution and show that only the $\Psi_4$ and $\Psi_2$ are non-zero. It is verified numerically, that all the coordinate free components of the Weyl and Ricci tensors decrease to arbitrary small values, and the metric tensor approaches a constant value in the mean. It is in this sense that the solution is said to asymptote Minkowski space. Section \ref{results} is devoted to a numerical extraction of the oscillatory part of the metric during the isotropisation process, which is interpreted as a combination of a tensor and a scalar perturbations. In section \ref{linearized f.eq.} we work out the perturbative scheme for the field equations, and present the solution of the first order metric perturbations for the Bianchi $I$ case studied here. Our conclusions and final remarks are presented in section \ref{conclusions}.

The following conventions and units are taken
$R_{bcd}^{a}=\Gamma_{bd,c}^{a}-...$, $R_{ab}=R_{acb}^{c}$,
$R=R_{a}^{a}$, metric signature $-+++$, the letters $a,b,...$ run from
$0-3$, while $i,j,...$ run from $1-3$ and $G=\hbar=c=1$.

\section{The field equations}\label{The field equations}

The Lagrangian density we consider here is
\begin{equation}\label{acao}
{\cal L} = \sqrt{-g}\left[\Lambda+R+\alpha\left(R^{ab}R_{ab}
-\frac{1}{3}R^{2}\right)+\beta R^{2}\right]+\mathcal{L}_{q}\, ,
\end{equation}
where $\mathcal{L}_q$ is the quantum part of the Lagrangian, $\alpha$ and $\beta$ are constants.
Notice that the validity of (\ref{acao}) as a classical effective theory is related to the following
conditions: the geometry fluctuates in a much larger spatial scale and much slower time scale
than those related to the quantum fields in $\mathcal{L}_q$. In a different context, a question could be raised if the gravitational field itself should be considered as an effective theory in a broader quantum gravity scenario. Although at the first order perturbation the classical and quantum theory are the same, we stress that in this present work, we assume that the gravitational field as a strict classical field.

By varying the action $I = \int  {\cal L} d^4x$ with respect to the metric we find the field equations
\begin{equation}\label{tensor E}
E_{ab}=G_{ab}+\left(\beta-\frac{1}{3}\alpha\right)H_{ab}^{(1)}+\alpha
H_{ab}^{(2)}-\langle T_{ab} \rangle-\frac{1}{2}\Lambda g_{ab}\equiv 0,
\end{equation}
where $\langle T_{ab} \rangle$ is the renormalized vacuum expectation value of the energy momentum tensor and
\begin{eqnarray}
 && H_{ab}^{(1)}=\frac{1}{2}g_{ab}R^{2}-2RR_{ab}+2R_{;ab}-2\square Rg_{ab},\label{tensor H1}\\
 && H_{ab}^{(2)}=\frac{1}{2}g_{ab}R_{mn}R^{mn}+R_{;ab}-2R^{cn}R_{cbna}-\square R_{ab}-\frac{1}{2}\square Rg_{ab},\label{tensor H2}\\
 && G_{ab}=R_{ab}-\frac{1}{2}g_{ab}R.\label{tensor G}
\end{eqnarray}

The counterterms $R^2$, $R_{ab}R^{ab}$, $\Lambda$  and $R$ in (\ref{acao}) are precisely the ones necessary to obtain a finite vacuum expectation  value of the energy momentum tensor (see for example \cite{christensen}). A theory without these counterterms is inconsistent from the point of view of the renormalization of the quantum field. Here, the renormalized vacuum expectation value of the energy momentum tensor is set to zero and we are disregarding any classical contribution to the energy momentum tensor.

The following Bianchi Type $I$ line element is considered
\begin{equation}
ds^{2}=-dt^{2}+[a_{1}(t)]^{2}dx^{2}
+[a_{2}(t)]^{2}dy^{2}+[a_{3}(t)]^{2}dz^{2},\label{elinha}
\end{equation}
which is spatially flat and anisotropic, and with $t$ being the proper time. The substitution of the above line element into $E_{11}\equiv0, \; E_{22}\equiv0, \;E_{33}\equiv0 $ of (\ref{tensor E}) with  $\langle T_{ab} \rangle=0$, yields  a set of ordinary differential equations of the type
\begin{eqnarray}
 &  & \frac{d^{4}}{dt^{4}}a_{1}=f_{1}\left(\dddot{a}_{i},\ddot{a}_{i},\dot{a}_{i},a_{i}\right)\label{edo1}\\
 &  & \frac{d^{4}}{dt^{4}}a_{2}=f_{2}\left(\dddot{a}_{i},\ddot{a}_{i},\dot{a}_{i},a_{i}\right)\label{edo2}\\
 &  & \frac{d^{4}}{dt^{4}}a_{3}=f_{3}\left(\dddot{a}_{i},\ddot{a}_{i},\dot{a}_{i},a_{i}\right),\label{edo3}
\end{eqnarray}
where the functions $f_{i}$ involve the $a_{1},\; a_{2},\; a_{3}$ and their derivatives, see  \ref{ap1}.

In addition to eqs. (\ref{edo1})-(\ref{edo3}), we have the
temporal part of eq. (\ref{tensor E}). To understand the role of this
equation we have first to study the covariant divergence of eq. (\ref{tensor E}), namely
\begin{equation}
\nabla_{a}E^{ab}=\partial_{a}E^{ab}+\Gamma_{ac}^{a}E^{cb}+\Gamma_{ac}^{b}E^{ac}=0.
\end{equation}

The temporal part of the above equation reads

\begin{equation}
\partial_{0}E^{00}+\Gamma_{a0}^{a}E^{00}+\Gamma_{00}^{0}E^{00}=0,
\label{vinculo}
\end{equation}
where $E_{00}$ is the $00$ component of eq. (\ref{tensor E}) From eq. (\ref{vinculo})we conclude that
if initially $E_{00}=0$, it will remain zero at any instant. Therefore the
equation for $E_{00}$ acts as a constraint on the initial conditions and we use it to test the accuracy of our results. Notice that the full numerical solution of this problem was worked out in Refs. \cite{sandro,daniel_sandro}.

\subsection{Perturbative approximation \label{sec.perturb}}

Following MTW \cite{MTW} we consider metric perturbations in the form
\begin{eqnarray}
 &  & g_{ab}=g_{ab}^{(B)}+\varepsilon h_{ab}\\
 &  & g^{ab}=g^{ab}_{(B)}-\varepsilon h^{ab}+\varepsilon^{2}h^{ac}h_{c}^{\; b}+...
\end{eqnarray}

In a free falling frame we have $g_{ab}^{(B)}=\eta_{ab}$ for the
background metric, and locally ${\Gamma^{(B)}}^a_{bc}=0$, so that
the connection compatible with $g_{ab}$ is identical to
$\Gamma_{ab}^{c}=\varepsilon S_{\; ab}^{c}$. Therefore, we have
\begin{eqnarray}
 &  & \nabla_{a}T_{b}=T_{a;b}=T_{a|b}-\varepsilon S_{\; ab}^{c}T_{c} \\
 &  & S_{ab}^{c}=\frac{1}{2}g_{(B)}^{cs}\left(h_{sa|b}+h_{sb|a}-h_{ab|s}\right),\label{conexao}
\end{eqnarray}
where $|$ is the covariant derivative with respect to the
background metric $g_{ab}^{(B)}$ and $\nabla_{a}=;a$ is the
covariant derivative with respect to $g_{ab}$. The tensors that
appear in eqs. (\ref{tensor E})-(\ref{tensor G}) can
be expanded in powers of $\varepsilon$, and to first order became equal to
\begin{eqnarray}
 &  & G_{ab}^{1}=R_{ab}^{1}-\frac{1}{2}h_{ab}R^{(B)}-\frac{1}{2}g_{ab}^{(B)}R^{1}\nonumber\\
 &  & H_{ab}^{(1)1}=2R_{|ab}^{1}-2\square^{(B)}R^{1}g_{ab}^{(B)}\nonumber\\
 &  & H_{ab}^{(2)1}=R_{|ab}^{1}-\square^{(B)}R_{ab}^{1}
 -\frac{1}{2}\square^{(B)}R^{1}g_{ab}^{(B)}.\label{ecl}
\end{eqnarray}
where
$R^1$, $R^1_{ab}$,  are the Ricci scalar and the Ricci tensor to first order in
$\varepsilon$, 
\begin{eqnarray*}
&&R_{ab}^1=\frac{1}{2}\left\{h^n_{a|bn}+h^n_{b|an}-\square h_{ab}-h^n_{n|ab}\right\}\\
&&R^1=\frac{1}{2}\left\{h^{mn}_{|mn}-\square h_n^n \right\}
\end{eqnarray*}
and $R_{abcd}^{1}$ is the Riemann tensor to first order (see \cite{MTW}) and $\square^{B}T^{a}=g^{Bmn}\nabla_{m}^{B}\nabla_{n}^{B}T^{a}=T_{\,|c}^{a\,\,|c}$.

The combination of the above equations gives a perturbative sum for
$E_{ab}$, namely
\begin{eqnarray}
 &  & E_{ab}=E_{ab}^{(B)}+\varepsilon E_{ab}^{1}+..\nonumber \\
 &  & E_{ab}^{1}=G_{ab}^{1}+\left(\beta-\frac{1}{3}\alpha\right)H_{ab}^{(1)1}+\alpha H_{ab}^{(2)1}-\frac{1}{2}\Lambda h_{ab}\label{tensor E1}
 \end{eqnarray}
where $E_{ab}^{(B)}$ correspond to the background equations satisfying $E_{ab}^{(B)}\equiv0.$
The linearized partial differential equations for the perturbation, $h_{ab}$, which are shown
in the  \ref{ap2}, are obtained by imposing $E_{ab}^{1}=0$.

\subsection{Instabilities \label{tachyon}}

Before proceeding, it is worth considering the issue concerning the
condition for the existence or not of tachyonic modes and ghosts. A concise reproduction of these well known results is presented in \ref{ap3} and we indicate Stelle's article \cite{Stelle1978} for further details and a deeper analysis. 

According to \cite{Stelle1978}, the linearized field equations have eight degrees of freedom. Two correspond to the familiar massless spin 2 graviton. Five correspond to a massive spin 2 particle with mass $m_2=1/{\sqrt{\alpha}}$. The last degree of freedom correspond to a massive scalar particle  with mass $m_0=1/{\sqrt{-6\beta}}$. If $\alpha>0$ and $\beta<0$ the tachyonic mode is excluded. The presence of the tachyonic mode in our case, would indicate the linear instability of the solution. In the present letter we choose $\alpha=2$ and $\beta=-1$ and therefore there is no tachyonic mode; these same values of $\alpha$ and $\beta$ were considered in the reference \cite{sandro,daniel_sandro}. 

The theory we are considering is not coupled to other fields in the sense that the eight degrees of freedom mentioned above are entirely free in a first order perturbative scheme. In this sense, only in the non linear regime there will be coupling between these modes and an infinite amount of energy can be extracted from the ghost field to the other linearized fields, which also induces instabilities, as can be seen for example in \cite{sandro,daniel_sandro}. 

\section{The asymptotic solution}\label{Weyl}

\begin{figure}
\resizebox{\imsize}{!}{\includegraphics{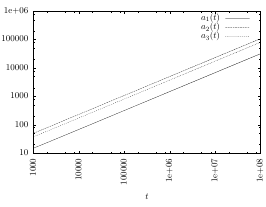}}
\caption{Numerical solution of eqs. (\ref{edo1}), (\ref{edo2}) and (\ref{edo3}) given in (\ref{A10}) for $t>1000$ plotted in a log-log scale. The initial condition is given in table \ref{tabela} in  \ref{ap2}. In this plot it is possible to obtain the mean time evolution of the scale factors to be, $a_1(t)=\alpha_1t^{\gamma_1}\simeq 0.14 t^{0.66}$, $a_2(t)=\alpha_2t^{\gamma_2}\simeq 0.50 t^{0.66}$, $a_3(t)=\alpha_3t^{\gamma_3}\simeq 0.38 t^{0.66}$, see table \ref{tabela} in  \ref{ap2}. Thus, the derivatives of the scale factors decrease with time to arbitrary small values, in the mean. \label{fig1_1}} 
\end{figure}

In Figure \ref{fig1_1} we show a particular complete, non perturbed, numerical solution of eqs. (\ref{edo1}), (\ref{edo2}) and (\ref{edo3}) given in (\ref{A10})
\begin{equation}
g_{ab}(t) = \left(\begin{array}{cccc}
-1 & 0 & 0 & 0\\
0 &a_1(t)^2 & 0 & 0\\
0 & 0 &a_2(t)^2  & 0\\
0 & 0 & 0 & a_3(t)^2\end{array}\right),
\label{metrica_g}
\end{equation} 
for $\alpha = 2$, $\beta = -1$ and $\Lambda =
0$. The initial condition in Figure \ref{fig1_1} is given by $\mbox{\ensuremath{\dot{a}}}_{1}=0.1,\,\dot{a}_{2}=0.3,\,a_{1}=1,\, a_{2}=1,\, a_{3}=1$, and $\dot{a}_{3}$ is chosen in
accordance with the temporal part $E_{00}=0$ of the field equation eq. (\ref{tensor E}). Remind that $E_{00}$ acts as a hamiltonian constraint: if it is zero initially it remains zero at any instant. All other higher order derivatives are chosen to be zero. 

We stress that according to Figure \ref{fig1_1}, the mean time evolution of the scale factors
\begin{eqnarray}
&&\langle a_1\rangle=\alpha_1t^{\gamma_1}\simeq 0.14 t^{0.66}\nonumber\\
&&\langle a_2\rangle =\alpha_2t^{\gamma_2}\simeq 0.50 t^{0.66}\nonumber\\
&&\langle a_3\rangle =\alpha_3t^{\gamma_3}\simeq 0.38 t^{0.66},\label{ev.media}
\end{eqnarray}
see table \ref{tabela} in  \ref{ap2}, are such that the time derivatives of the scale factors always decrease in time. It can be seen that the exponents are almost the same. We intend to investigate this point further elsewhere. According to the line element in (\ref{elinha}), for $t\rightarrow \infty$ this should be enough to state that asymptotically the solution approaches Minkowski space. Anyway, we proceed to show numerically that the Riemann tensor vanishes in a coordinate invariant fashion. 

A complex null basis can be defined 
\begin{eqnarray}
&&k^a=1/\sqrt{2}(1,1/a_1(t),0,0)\nonumber\\ 
&&l^a=1/\sqrt{2}(1,-1/a_1(t),0,0)\nonumber\\
&&t^a=1/\sqrt{2}(0,0,1/a_2(t),-i/a_3(t))\nonumber\\
&&\bar{t}^a=1/\sqrt{2}(0,0,1/a_2(t),i/a_3(t))\nonumber\\
&&k^{a}k_{a}=k^{a}t_{a}=k^{a}\bar{t}_{a}=l^{a}l_{a}=l^{a}t_{a}=l^{a}\bar{t}_{a}=t^{a}t_{a}=\bar{t}^{a}\bar{t}_{a}=0,\nonumber\\
&&t^{a}\bar{t}_{a}=-k^{a}l_{a}=1,
\label{basenula}
\end{eqnarray}
with the corresponding null metric 
\[
\tilde{g}_{AB}=g_{ab}A^aB^b=\left(\begin{array}{cccc}
0&-1 & 0 & 0\\
-1& 0 & 0&0\\
0 & 0&0&1\\
0&0 &1&0
\end{array}\right),
\]
where $A^a$ and $B^b$ are the null vectors in (\ref{basenula}). The Newman-Penrose complex coefficients are in fact the tetrad components of the 
Weyl tensor $C_{abcd}$
\begin{eqnarray*}
&&\Psi_{0}=C_{abcd}k^{a}t^{b}k^{c}t^{d},\\
&&\Psi_{1}=C_{abcd}k^{a}l^{b}k^{c}t^{d},\\
&&\Psi_{2}=C_{abcd}k^{a}t^{b}\bar{t}^{c}l^{d},\\
&&\Psi_{3}=C_{abcd}k^{a}l^{b}\bar{t}^{c}l^{d},\\
&&\Psi_{4}=C_{abcd}\bar{t}^{a}l^{b}\bar{t}^{c}l^{d}.
\end{eqnarray*}

\begin{figure}
\begin{center}
\resizebox{\imsize}{!}{\includegraphics{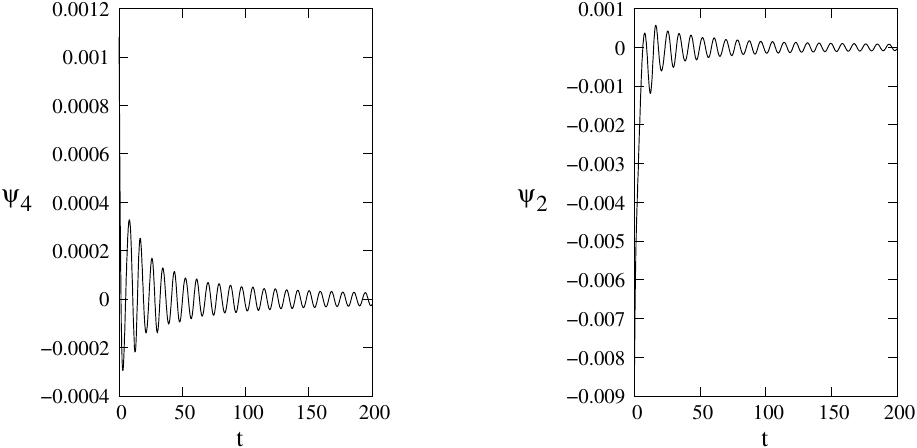} }
\end{center}
\caption{Behavior of the temporal decay of the Weyl's invariants $\Psi_4$  and $\Psi_2$ for the non perturbed solution for $a_1(t)$, $a_2(t)$ and $a_3(t)$ shown in Figure \ref{fig1_1}.  It can be seen that the oscillating behaviour of the Weyl scalars is in good agreement with eq. (\ref{frequencia}), namely $\omega=1/\sqrt{\alpha}=1/\sqrt{2}$.\label{Psi_4}}
\end{figure}

For the particular line element given by (\ref{metrica_g}) the Weyl scalars read
\begin{eqnarray*}
&&\Psi_4\equiv\Psi_0=-\frac{1}{4}\left( \frac{\dot{a}_1}{a_1}\frac{\dot{a}_3}{a_3}-\frac{\dot{a_1}}{a_1}\frac{\dot{a}_2}{a_2}+\frac{\ddot{a}_2}{a_2}-\frac{\ddot{a}_3}{a_3}\right)\\
&&\Psi_2=\frac{1}{12}\left(-2\frac{\dot{a}_2}{a_2}\frac{\dot{a}_3}{a_3}+
\frac{\dot{a}_1}{a_1}\frac{\dot{a}_3}{a_3} +\frac{\dot{a}_1}{a_1}\frac{\dot{a}_2}{a_2}
+\frac{\ddot{a}_2}{a_2} +\frac{\ddot{a}_3}{a_3}-2\frac{\ddot{a}_1}{a_1}\right)\\
&&\Psi_1\equiv \Psi_3\equiv 0,
\end{eqnarray*}
which are shown in Figure \ref{Psi_4}. They were evaluated for the initial conditions and parameters considered above
and, albeit being non zero, as can be seen in Figure \ref{Psi_4}, all the $\Psi$ decrease in time to arbitrary small values.

\begin{figure}
\resizebox{\imsize}{!}{\includegraphics{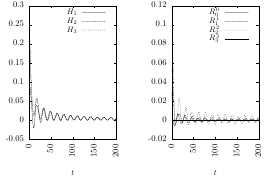}}
\caption{For the same initial condition and parameters of Figure \ref{fig1_1} we show the numerical evolution of  the  Hubble constants in each direction $H_{1}=\dot{a}_1/a_1,$ $H_2=\dot{a}_2/a_2$ and $H_3=\dot{a}_3/a_3$ and the Ricci tensor $R^a_b$, according of the non linearized differential equations  (\ref{edo1}), (\ref{edo2}) and (\ref{edo3}). \label{Figura_u}}
\end{figure}

Again according to the line element (\ref{metrica_g}), the tetrad components of the Ricci tensor  
\begin{eqnarray*} 
&&R_0^{\;0}=-\frac{\ddot{a}_1}{a_1}-\frac{\ddot{a}_2}{a_2}-\frac{\ddot{a}_3}{a_3}\\
&&R_i^{\;j}=-\left(\begin{array}{ccc}
\frac{\ddot{a}_1}{a_1}+\frac{\dot{a}_1}{a_1}\frac{\dot{a}_3}{a_3}+\frac{\dot{a}_1}{a_1}\frac{\dot{a}_2}{a_2} & 0 & 0\\
0 & \frac{\ddot{a}_2}{a_2}+\frac{\dot{a}_2}{a_2}\frac{\dot{a}_3}{a_3}+\frac{\dot{a}_2}{a_2}\frac{\dot{a}_1}{a_1} & 0\\
0 & 0 & \frac{\ddot{a}_3}{a_3}+\frac{\dot{a}_3}{a_3}\frac{\dot{a}_1}{a_1}+\frac{\dot{a}_3}{a_3}\frac{\dot{a}_2}{a_2}\end{array}\right),
\end{eqnarray*}
and the Hubble constants in each direction $H_i=\dot{a}_i/a_i$, have a decreasing behavior, as  can be seen in Figure \ref{Figura_u}. 

The  Riemann tensor can be written as
\[
R_{abcd}=C_{abcd} +\frac{1}{2}(g_{ca}R_{bd}+g_{db}R_{ca}-g_{cb}R_{da}-g_{da}R_{cb})\\
-\frac{1}{6}R(g_{ca}g_{db}-g_{cb}g_{da}),
\]
where $C_{abcd}$ is the Weyl tensor. According to Figures \ref{Psi_4} and \ref{Figura_u} the tetrad components of the Weyl and Ricci tensor always decrease, $C_{abcd}\rightarrow 0$ and $R_{ab}\rightarrow 0$. Then the coordinate free components of the Riemann tensor also vanish asymptotically
\[R_{abcd}\rightarrow 0.\] 

Since the Riemann tensor vanishes and the mean value of the time derivatives of the $a_i(t)$ have numerically decreased to arbitrary small values, as can be seen in (\ref{ev.media}) and in Figures \ref{fig1_1} and \ref{fig1_1} this solution is said to asymptote Minkowski space. It shows isotropisation in the strong sense.

One interesting issue is to see how the oscillating behaviors of $\Psi_4$ and $\Psi_2$ depend
on the parameters $\alpha$ and $\beta$. To this aim one can consider that
\[g_{ab}=\eta_{ab} + h_{ab}e^{i\omega t},\]
with $|\dot{h}_{ab}|<<\omega$. Substituting the above expression into the linearized equations, which are shown in the Appendix \ref{ap2}, one has the following eigenvalues
\begin{equation}
\omega=0,\;\;\omega=\frac{1}{\sqrt{\alpha}},\;\;\omega=\frac{1}{\sqrt{-6\beta}},
\label{frequencia}
\end{equation}
in Riemann local coordinates, where $|h_{ab}|<<1$. In our case, the oscillating behavior of the Weyl scalars is in good agreement with $\omega = 1/\sqrt{2}$ as can be checked in Figure \ref{Psi_4}.  Of course the frequencies $\omega$ in (\ref{frequencia}), agree with the masses (\ref{m0}) and (\ref{m2}) and the massless spin-2 particle $\phi_{ab}$ in (\ref{campos220}). We stress that (\ref{frequencia}) is in fact the spectrum for the strict spatially homogenous Bianchi $I$ case. The purpose of this present letter is to understand the solution itself instead of introducing a departure from the exact Bianchi $I$ case through the spatial dependence of the perturbation, which would give a dispersion relation for $\vec{k}\neq 0$. As is well known, for massive particles it is possible that $\omega\neq 0$ and $\vec{k}=0$. And in this case remind that the linearized fields can be understood as a massive spin 2 particle and a massive spin 0 particle. 

\section{Numerical results: extracting the metric perturbations}\label{results}

In this section we consider the issue of how to extract the metric perturbations from the above numerical solution for large $t$.

In the following two sections, we consider an arbitrary time interval, say, around $t=t_0\simeq 8 \times 10^7$ in natural units for which the spacetime is close to Minkowski space in the above sense. Applying a linear regression to this interval we find, according to (\ref{ev.media}) the following mean expression for the three scale factors
\begin{eqnarray*} 
 &&\langle a_{1}\rangle\simeq \alpha_1(1-\gamma_1)t_0^{\gamma_1}+\alpha_1\gamma_1t_0^{\gamma_1-1}t=c_{1}+b_{1}t\\
 && \langle a_{2}\rangle\simeq\alpha_2(1-\gamma_2)t_0^{\gamma_2}+\alpha_2\gamma_2t_0^{\gamma_2-1}t=c_{2}+b_{2}t\\
 && \langle a_{3}\rangle\simeq \alpha_3(1-\gamma_3)t_0^{\gamma_3}+\alpha_3\gamma_3t_0^{\gamma_3-1}t=c_{3}+b_{3}t, 
\end{eqnarray*}
which by convenience we rewrite it as  
\begin{eqnarray}
\label{regresslinear}
\langle g_{ab}\rangle &=& \left(\begin{array}{cccc}
-1 & 0 & 0 & 0\\
0 &(c_1+b_1t)^2 & 0 & 0\\
0 & 0 &(c_2+b_2t)^2  & 0\\
0 & 0 & 0 & (c_3+b_3t)^2\end{array}\right)\nonumber\\
&=&\left(\begin{array}{cccc}
-1 & 0 & 0 & 0\\
0 &c_1^2 & 0 & 0\\
0 & 0 &c_2^2  & 0\\
0 & 0 & 0 & c_3^2\end{array}\right)\nonumber\\
&&+
\left(\begin{array}{cccc}
0 & 0 & 0 & 0\\
0 &2c_1b_1t+(b_1t)^2 & 0 & 0\\
0 & 0 &2c_2b_2t+(b_2t)^2  & 0\\
0 & 0 & 0 & 2c_3b_3t+(b_3t)^2\end{array}\right)
\end{eqnarray}
where $b_1=2.29\times 10^{-4},\,b_2=7.79\times 10^{-4},\,b_3=5.82\times 10^{-4},\,c_1=9.17\times 10^3,\,c_2=31.2\times 10 ^3,\, {\rm and} \; c_3=23.3\times 10^3$, see table \ref{tabela} in the  \ref{ap2}. As already mentioned, we must stress that the coefficients $b_1$ $b_2$ and $b_3$, which are the mean values of the time derivatives of the scale factors, decrease to arbitrary small values, see Figure \ref{fig1_1}. In the following we intentionally keep the linear regression and include it into the perturbation of the metric. The only reason for this is to specifically compare the oscillating part of the metric obtained by the two different methods, see section \ref{linearized f.eq.}.

\begin{figure}
\resizebox{\imsize}{!}{\includegraphics{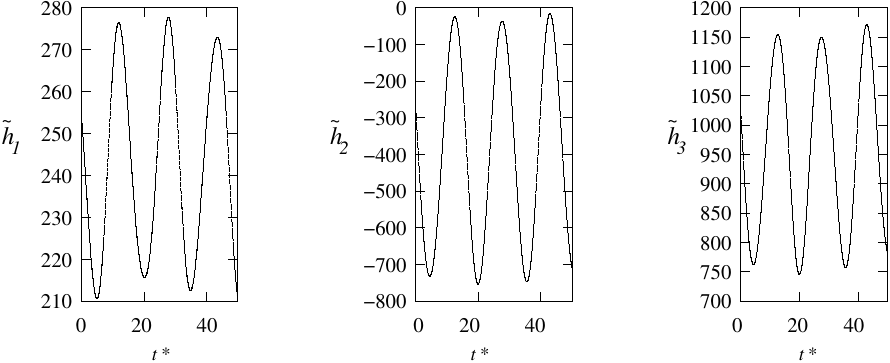} }
\caption{The oscilating part of
$\tilde{h}_{i}(t)=[a_{i}(t)]^{2}-[c_{i}+b_{i}t]^{2}$ as a function of time.
The values of $a_{i}$ are the numerical solutions of eqs. (\ref{edo1})-(\ref{edo3}) as shown in Figure \ref{fig1_1}.
The values of $c_{i}$ and $b_{i}$ are given in table \ref{tabela} of the  \ref{ap2}. This plot together with Figure \ref{fig1_1}, shows that the solution has a slow time evolution superposed with much faster time oscillating functions $\tilde{h}_i(t)$. We stress that the amplitude of the $\tilde{h}_i(t)$ is much smaller than the mean time evolution of the scale factors shown in Figure \ref{fig1_1}. Note that the time origin was changed to $t=8.0001\times 10^7 +t^*$. \label{h(t)}}
\end{figure}

Now, subtracting eq. (\ref{regresslinear}) from the complete numerical solution of eqs. (\ref{edo1}), (\ref{edo2}) and (\ref{edo3}), we find that a small oscillatory part of the metric remains
\[
g_{ab}(t)-\langle g_{ab}\rangle=\left(\begin{array}{cccc}
0 & 0 & 0 & 0\\
0 & \tilde{h}_{1}(t) & 0 & 0\\
0 & 0 & \tilde{h}_{2}(t) & 0\\
0 & 0 & 0 & \tilde{h}_{3}(t)\end{array}\right),
\]
which is shown in Figure \ref{h(t)}. Thus, rearranging the metric of the complete solution we find that it can be written in the following way 
\begin{equation}
g_{ab} \doteq g^{(B)}_{ab} + \tilde{h}_{ab},
\end{equation}
where $g^{(B)}_{ab}$ is a constant background given by
\[g^{(B)}_{ab} = {\rm diag}(-1,c_1^2, c_2^2, c_3^2) \]
and ${\tilde{h}}_{ab}$ is a superposition of three time oscillating
functions $\tilde{h}_1(t)$, $\tilde{h}_2(t)$ and $\tilde{h}_3(t)$
with a linear and quadratic term in time, namely

\begin{eqnarray}
\label{h tilde}
&&\tilde{h}_{ab}(t) = \nonumber\\
&&\left(\begin{array}{cccc}
0 & 0 & 0 & 0\\
0 & \tilde{h}_{1}(t)+2c_{1}b_{1}t+b_{1}^{2}t^{2} & 0 & 0\\
0 & 0 & \tilde{h}_{2}(t)+2c_{2}b_{2}t+b_{2}^{2}t^{2} & 0\\
0 & 0 & 0 & \tilde{h}_{3}(t)+2c_{3}b_{3}t+b_{3}^{2}t^{2}\end{array}\right).
\end{eqnarray}

Note that the sum $g_{ab}^{(B)} + \tilde{h}_{ab}$ is identical to the metric 
\[g_{ab} = {\rm diag}(-1, a_1(t)^2, a_2(t)^2, a_3(t)^2),\] 
which is the non perturbed solution of (\ref{tensor E})shown in Figure \ref{fig1_1}, in the time interval $t\simeq 8\times 10^7$.

\begin{figure}
\resizebox{\imsize}{!}{\includegraphics{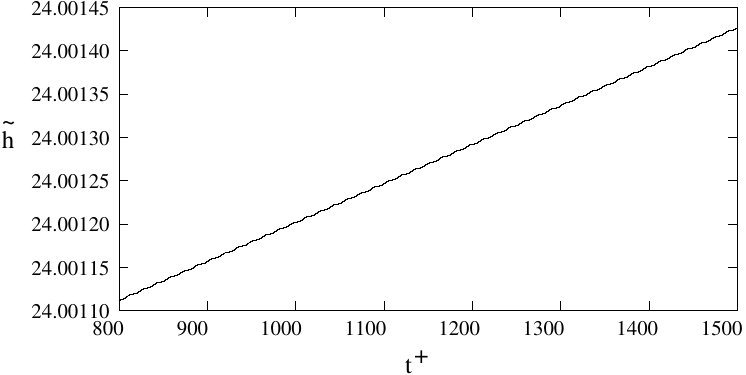}}
\caption{Spatial trace $\tilde{h}=g^{(B)\,ij}\tilde{h}_{ij}$ as a function of time.
It can be seen that the trace of $\tilde{h}$ is almost constant
in comparison with the individual amplitudes of the components of $\tilde{h}_{ab}$.
Here, a new time origin was chosen to be
$t=8.0\times 10^7+ t^+$.\label{traco h}}
\end{figure}

Now, we need to ask if the metric perturbation
$\tilde{h}_{ab}$ indeed represents the tensor contribution. We see that $\tilde{h}_{ab}$ is
spatially divergenceless (as it depends only on time), its time components are null (as can be seen from eq.
(\ref{h tilde}), but it is not traceless. Thus, in order to
obtain the tensorial component we need to subtract
the trace of $\tilde{h}_{ab}$ generating a new quantity, namely
\begin{equation}
h_{ab} = \tilde{h}_{ab}  - \frac{1}{4}g^{(B)}_{ab}\tilde{h},
\end{equation}
which finally has the following properties
\begin{equation}
{h^{ab}}_{|b} = h^{a0} = h = 0,
\end{equation}
where the symbol $|$ represents the covariant derivative with respect to the background metric $g_{(B)}^{ab}$. The quantities $h_{ab}$ are nothing but the TT (transverse-traceless) components of $\tilde{h}_{ab}$, while the trace $\tilde{h} = g_{(B)}^{ab}\tilde{h}_{ab}$ is the scalar perturbation shown in Figure \ref{traco h} that varies linearly with time but with a very small time derivative making it almost constant in the time interval considered.

The above procedure shows that during the isotropisation process the spacetime in quadratic gravity can be described as the superposition of the Minkowski metric with scalar and tensor gravitational waves. The gravitational perturbations are defined as a combination of an oscillatory part and a term that increases weakly with time.

\section{Solving the linearized field equations}\label{linearized f.eq.}

A second and straightforward method to investigate the evolution of perturbations that come from the isotropisation process is to write down explicitly the field equations for the first order small perturbations in the asymptotic region. This section in fact, is a demonstration that the solution is a perturbation over Minkowski space $g^{(B)}_{ab}={\rm diag}[-1,c_1^2,c_2^2,c_3^2]$.

\subsection{The linearized solution}

Now, let us consider the background metric as $g^{(B)}_{ab}={\rm diag}[-1,c_1^2,c_2^2,c_3^2]$ with the $c_i$ constants given in table \ref{tabela} in the  \ref{ap2}. With this background metric we have $E^{(B)}_{ab}\equiv0$, and the linear system of equations to be solved is obtained by setting $E_{ab}^{1}\equiv0,$ resulting in

\begin{eqnarray}\label{equacao de movimento linear}
 &  & \frac{d^{4}}{dt^{4}}\hat{h}_{1}=f_{1j}^{3}\left(t\right)\dddot{\hat{h}}_{j}+f_{1j}^{2}(t)\ddot{\hat{h}}_{j}+f_{1j}^{1}(t)\dot{\hat{h}}_{j}+f_{1j}^{0}(t)\hat{h}_{j}\nonumber \\
 &  & \frac{d^{4}}{dt^{4}}\hat{h}_{2}=f_{2j}^{3}\left(t\right)\dddot{\hat{h}}_{j}+f_{2j}^{2}(t)\ddot{\hat{h}}_{j}+f_{2j}^{1}(t)\dot{\hat{h}}_{j}+f_{2j}^{0}(t)\hat{h}_{j} \\
 &  & \frac{d^{4}}{dt^{4}}\hat{h}_{3}=f_{3j}^{3}\left(t\right)\dddot{\hat{h}}_{j}+f_{3j}^{2}(t)\ddot{\hat{h}}_{j}+f_{3j}^{1}(t)\dot{\hat{h}}_{j}+f_{3j}^{0}(t)\hat{h}_{j}\nonumber
\end{eqnarray}
where $\hat{h}$ is defined as the oscillatory part of the metric
perturbation and the coefficients $f_{ij}^{a}(t)$ also depend on
the parameters $\alpha,\,\beta,$ and $\Lambda$, as can be seen in the  \ref{ap2}.

\begin{figure}
\resizebox{\imsize}{!}{\includegraphics{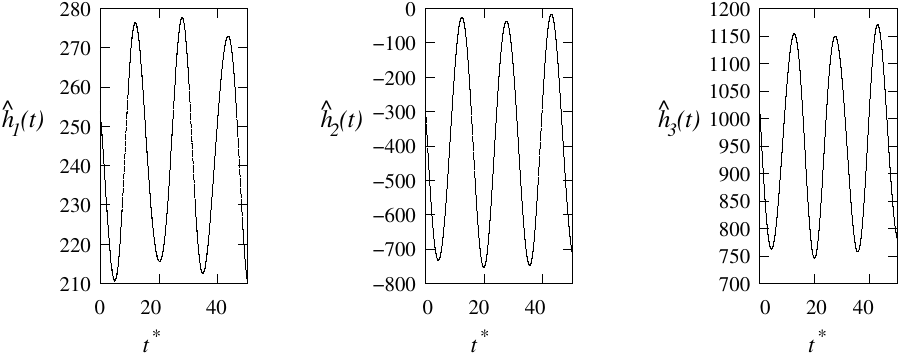}}
\caption{The time evolution of $\hat{h}_{i}$ for the linearized
system given in eqs. (\ref{equacao de movimento linear}) for an initial condition
consistent with the one shown in Figure \ref{h(t)}.
It can be seen that the solution is indistinguishable from Figure \ref{h(t)}. Again, note that the time origin was changed to $t=8.0001\times 10^7 +t^*$.\label{h_l}}
\end{figure}

The initial condition for the above set of differential equations is chosen to be consistent
with the solution shown in Figure \ref{fig1_1} and is shown in table \ref{tabela} of the  \ref{ap2}. Again, the values of $\alpha=2.0$, $\beta=-1.0$ and $\Lambda=0.0$, as in the preceding sections. In Figure \ref{h_l} we present the numerical solution of the above set of linear differential equations. Notice
that Figures \ref{h_l} ($\hat{h}_{i}(t)$) and \ref{h(t)} ($\tilde{h}_{i}(t)$) are identical. Also according to Figures \ref{h_l} or \ref{h(t)} together with Figure \ref{fig1_1}, shows that the solution has a slow time evolution superposed with much faster time oscillating functions $\tilde{h}_i(t)$. We stress that the oscillating part of the solution is included in Figure \ref{fig1_1}, and is so small that it's almost impossible to be seen. Stated differently, the  amplitude of the $\tilde{h}_i(t)$ is much smaller than the mean time evolution of the scale factors shown in Figure \ref{fig1_1}, which is in agreement with the adopted perturbative approach.

\section{Final remarks}\label{conclusions}

In the present letter we considered general anisotropic
Bianchi $I$ homogeneous spacetimes with the line element given by eq.
(\ref{elinha}). For this line element, the theory given by eq.
(\ref{acao}) reduces to a system of forth order ordinary nonlinear
differential equations. This system was investigated numerically.
The renormalized vacuum expectation value
of the energy momentum tensor was set to zero and we disregarded
any classical contribution to the energy momentum tensor.

We found that at the end of the process of isotropisation within quadratic gravity,
the solution is a superposition of Minkowski spacetime with homogenous scalar and tensor
gravitational waves.

According to Stelle \cite{Stelle1978}, there are eight degrees of freedom in the linearized limit of quadratic gravity. Two correspond to the familiar massless spin 2 graviton. Five correspond to a massive spin 2 particle. The last degree of freedom correspond to a massive scalar particle. Our solution is in accordance with Stelle's article, as the tachyon was excluded by the numerical choice of the parameters entering the quadratic action (\ref{acao}) and the ghost field is a totally free field, unable to induce instabilities. 

There is a recent review on physical and observational consequences on a non homogenous universe coupled to matter fields, out of which a Bianchi $I$ phase emerges, \cite{prd}. The theory we are addressing results from vacuum polarization of a quantized field on a curved, classical background. In this sense, we are also investigating the coupling of gravity with other fields. In this approach, gravity is considered a classical, non quantized field, and the resulting effective action is not to sensitive to the details of the quantized field, being almost the same for scalar, vector or spinorial fields \cite{christensen}. Concerning gravitational waves, the most important would be a spectrum which is in fact given by (\ref{frequencia}). On the other hand, the purpose of this present letter is to understand the solution itself instead of introducing a departure from the strict Bianchi $I$ case through the spatial dependence of the perturbation. We believe that the generalized dispersion relation with $\vec{k}\neq 0$ could be studied in the future.

For the specific line element considered here, we found that the Weyl's invariants, $\Psi_1\equiv\Psi_3\equiv 0$ while  $\Psi_2$ and $\Psi_4\equiv \Psi_0$ are non-null for the non perturbed numerical solution, and decay to arbitrary small values. This behavior also occurs to the Ricci tensor, so that the solution asymptotes Minkowski space in the sense that the Riemann tensor, in tetrad coordinate free components, vanishes asymptotically. 

Then, we have adopted two distinct approaches in order to describe the linearized tensor $h_{ab}$. The two methods have coincided in the prediction of the oscillatory pattern of metric oscillations with a scalar, $h=g^B_{ab}h^{ab}$ and a tensor component$h_{ab}-g^B_{ab}h/4$. We conclude that the process of isotropisation given in this article, is a superposition of Minkowksi space with scalar and tensor gravitational waves as already predicted in Stelle's article \cite{Stelle1978}. 

%According to Bardeen's milestone work \cite{bardeen}, the tensorial component is an indication of the presence of gravitational waves. The modes of the perturbations are identified from the Helmholtz equation $\nabla ^2 \psi = -k^2 \psi$, where $\psi$ can be a scalar, a vector or a tensor perturbation.  Bardeen also discusses the homogenous mode ($k=0$), which for the spatially Euclidean $E^3$ case, does not have to depend on the spatial coordinates. For the spherical $S^3$ and hyperbolic $H^3$ spatial sections $k$ is respectively $k^2=3$ and $k^2=1$ (see also \cite{perturbacoes}). 

%On the other hand, Matarrese et al \cite{matarrese} argued that
%gravitational waves do not occur when the magnetic part of the Weyl tensor
%$H_{ab}=0$  vanishes. These authors are not concerned with the distinction
%between gravitational waves and gravitational radiation, in Isaacson's sense.
%Also, all Petrov type N solutions have a non vanishing $H_{ab}\neq
%0$ tensor. The spacetimes with vanishing $H_{ab}=0$, vanishing shear, vorticity
%and pressure were named silent universes, see for example \cite{dsc}.

%In fact, there are many definitions of gravitational waves, see for example \cite{zakharov}. In a broader (and perhaps non rigorous sense) many more solutions can be understood as gravitational waves. For instance, the exact Bianchi $VII_h$ Petrov type N solutions \cite{ogh} are homogeneous solutions. There is not a localized spatial source being the origin and there is not a spatially compact wave front involving the source.

Finally, it is worth mentioning that although we have restricted our study to the Bianchi $I$ spaces, it is expected that the results and conclusions obtained here should be valid for any Bianchi type, as long as the three-curvature of space becomes irrelevant, i.e. ${}^3R_{ijkl}\rightarrow 0$.

\section*{Acknowledgments}
D. M. wishes to thank the Brazilian projects: {\it Nova F\'\i sica no Espa\c co} and INCT-A. J.C.N.A. would like to thank CNPq and FAPESP for financial support. We thank Massimo Tinto for critically reading the paper.
%\newpage
\appendix
%dummy comment inserted by tex2lyx to ensure that this paragraph is not empty
%dummy comment inserted by tex2lyx to ensure that this paragraph is not empty
\section{Non perturbed field equations \label{ap1}}
The following line element is chosen 
\[
ds^2=-dt^2+e^{2w_{{1}}(t)}dx^2+e^{2w_{{2}}(t)}dy^2+e^{2w_{{3}}(t)}dz^2.
\]
With this choice, the first time derivatives of the functions
$w_{{i}}(t)$ are related to the functions $a_{{i}}(t)$ of (\ref{metrica_g}) in the
following manner
\begin{eqnarray*}
y_{{1}}&=&\dot{w_{{1}}}=\frac{\dot{a_{{1}}}}{a_{{1}}}\\        
y_{{2}}&=&\dot{w_{{2}}}=\frac{\dot{a_{{2}}}}{a_{{2}}}\\       
y_{{3}}&=&\dot{w_{{3}}}=\frac{\dot{a_{{3}}}}{a_{{3}}}\\ 
y_{{4}}&=&\ddot{w_{{1}}}=\frac{\ddot{a_{{1}}}}{a_{{1}}}-\left(\frac{\dot{a_{{1}}}}{a_{{1}}} \right)^2\\        
y_{{5}}&=&\ddot{w_{{2}}}=\frac{\ddot{a_{{2}}}}{a_{{2}}}-\left(\frac{\dot{a_{{2}}}}{a_{{2}}} \right)^2
\end{eqnarray*}
\begin{eqnarray*}          
y_{{6}}&=&\ddot{w_{{3}}}=\frac{\ddot{a_{{3}}}}{a_{{3}}}-\left(\frac{\dot{a_{{3}}}}{a_{{3}}} \right)^2\\                                 
y_{{7}}&=&\frac{d^3}{dt^3}w_{{1}}=\frac{1}{a_{{1}}}\frac{d^3}{dt^3}a_{{1}}-
\frac{\dot{a_{{1}}}}{a_{{1}}}\left(3\frac{\ddot{a_{{1}}}}{a_{{1}}}-2\left(\frac{\dot{a_{{1}}}}{a_{{1}}} \right)^2 \right)\\                           
y_{{8}}&=&\frac{d^3}{dt^3}w_{{2}}=\frac{1}{a_{{2}}}\frac{d^3}{dt^3}a_{{2}}-
\frac{\dot{a_{{2}}}}{a_{{2}}}\left(3\frac{\ddot{a_{{2}}}}{a_{{2}}}-2\left(\frac{\dot{a_{{2}}}}{a_{{2}}} \right)^2 \right)\\                               
y_{{9}}&=&\frac{d^3}{dt^3}w_{{3}}=\frac{1}{a_{{3}}}\frac{d^3}{dt^3}a_{{3}}-
\frac{\dot{a_{{3}}}}{a_{{3}}}\left(3\frac{\ddot{a_{{3}}}}{a_{{3}}}-2\left(\frac{\dot{a_{{3}}}}{a_{{3}}}
\right)^2 \right).
\end{eqnarray*}
The differential equations for the quadratic theory with the classical
source $\langle T_{ab}\rangle=0$, (\ref{tensor E}), are equivalent to 
\begin{eqnarray}
\frac{d^4}{dt^4}w_{{1}}&=&
{\frac {11}{9}}\,y_{{3}}y_{{1}}{y_{{2}}}^{2}-{\frac {19}{9}}\,y_{{3}}y
_{{2}}{y_{{1}}}^{2}+{\frac {11}{9}}\,y_{{2}}y_{{1}}{y_{{3}}}^{2}+5/3\,
y_{{1}}y_{{2}}y_{{5}}-13/3\,y_{{1}}y_{{2}}y_{{4}}\nonumber\\ 
&&-13/3\,y_{{1}}y_{{3}}
y_{{4}}+1/9\,y_{{1}}y_{{6}}y_{{2}}+5/3\,y_{{1}}y_{{3}}y_{{6}}+2/3\,y_{
{2}}y_{{3}}y_{{6}}-{\frac {14}{9}}\,y_{{2}}y_{{3}}y_{{4}}+2/3\,y_{{2}}
y_{{3}}y_{{5}}\nonumber\\ 
&&+1/9\,y_{{1}}y_{{3}}y_{{5}}- \left( {\frac {1}{108}}\,y_
{{3}}y_{{1}}{y_{{2}}}^{2}+{\frac {1}{108}}\,y_{{3}}y_{{2}}{y_{{1}}}^{2
}+{\frac {1}{108}}\,y_{{2}}y_{{1}}{y_{{3}}}^{2}+{\frac {1}{54}}\,y_{{1
}}y_{{6}}y_{{2}}\right.\nonumber\\
&&\left.+{\frac {1}{54}}\,y_{{2}}y_{{3}}y_{{4}}+{\frac {1}{54}
}\,y_{{1}}y_{{3}}y_{{5}}-{\frac {1}{108}}\,y_{{5}}y_{{4}}-{\frac {1}{
108}}\,y_{{6}}y_{{4}}-{\frac {1}{108}}\,y_{{6}}y_{{5}}+1/27\,{y_{{1}}}
^{2}{y_{{2}}}^{2}\right.\nonumber\\
&&\left.+1/27\,{y_{{1}}}^{2}{y_{{3}}}^{2}+1/27\,{y_{{2}}}^{2}
{y_{{3}}}^{2}-1/36\,{y_{{3}}}^{3}y_{{1}}-1/36\,{y_{{1}}}^{3}y_{{2}}-1/
36\,{y_{{1}}}^{3}y_{{3}}-1/36\,{y_{{2}}}^{3}y_{{1}}\right.\nonumber\\
&&\left.-1/36\,{y_{{3}}}^{3
}y_{{2}}-1/36\,{y_{{2}}}^{3}y_{{3}}-{\frac {1}{54}}\,y_{{1}}y_{{7}}-{
\frac {1}{54}}\,y_{{2}}y_{{8}}-{\frac {1}{54}}\,y_{{3}}y_{{9}}+{\frac 
{1}{108}}\,y_{{8}}y_{{1}}+{\frac
  {1}{108}}\,y_{{2}}y_{{7}}\right.\nonumber\\
&&\left.+{\frac {1}{
108}}\,y_{{9}}y_{{1}}+{\frac {1}{108}}\,y_{{9}}y_{{2}}+{\frac {1}{108}
}\,y_{{3}}y_{{8}}+{\frac {1}{108}}\,{y_{{4}}}^{2}+{\frac {1}{108}}\,{y
_{{5}}}^{2}+{\frac {1}{108}}\,{y_{{6}}}^{2}+{\frac {1}{108}}\,{y_{{2}}
}^{4}\right.\nonumber\\
&&\left.+{\frac {1}{108}}\,{y_{{3}}}^{4}-{\frac {1}{54}}\,{y_{{1}}}^{2}y_
{{4}}-{\frac {1}{54}}\,{y_{{2}}}^{2}y_{{5}}-{\frac {1}{54}}\,{y_{{3}}}
^{2}y_{{6}}+{\frac {1}{108}}\,y_{{3}}y_{{7}}+{\frac {1}{108}}\,{y_{{1}
}}^{4} \right) \alpha{\beta}^{-1} \nonumber\\
&&-\left( -1/18\,y_{{6}}-1/18\,y_{{5}}
+1/6\,\Lambda-1/18\,{y_{{3}}}^{2}-1/36\,y_{{3}}y_{{2}}-1/18\,{y_{{2}}
}^{2}-1/18\,y_{{4}}-1/36\,y_{{3}}y_{{1}}\right.\nonumber\\
&&\left.-1/18\,{y_{{1}}}^{2}-1/36\,y_{
{2}}y_{{1}}\right)/(G\beta)-{\frac {11}{9}}\,y_{{5}}y_{{4}}-{\frac {11}{9}}
\,y_{{6}}y_{{4}}+1/9\,y_{{6}}y_{{5}}-{\frac {5}{18}}\,{y_{{1}}}^{2}{y_
{{2}}}^{2}-{\frac {5}{18}}\,{y_{{1}}}^{2}{y_{{3}}}^{2}\nonumber\\
&&+{\frac {7}{18}}
\,{y_{{2}}}^{2}{y_{{3}}}^{2}+2/3\,{y_{{3}}}^{3}y_{{1}}-2/3\,{y_{{1}}}^
{3}y_{{2}}-2/3\,{y_{{1}}}^{3}y_{{3}}+2/3\,{y_{{2}}}^{3}y_{{1}}-{\frac 
{22}{9}}\,y_{{1}}y_{{7}}-1/9\,y_{{2}}y_{{8}}\nonumber\\
&&-1/9\,y_{{3}}y_{{9}}-1/9\,
y_{{8}}y_{{1}}-{\frac {16}{9}}\,y_{{2}}y_{{7}}-1/9\,y_{{9}}y_{{1}}+2/9
\,y_{{9}}y_{{2}}+2/9\,y_{{3}}y_{{8}}-{\frac {29}{18}}\,{y_{{4}}}^{2}
\nonumber\\
&&-{
\frac {5}{18}}\,{y_{{5}}}^{2}-{\frac {5}{18}}\,{y_{{6}}}^{2}-{\frac {5
}{18}}\,{y_{{2}}}^{4}-{\frac {5}{18}}\,{y_{{3}}}^{4}-1/9\,{y_{{1}}}^{2
}y_{{4}}-{\frac {7}{9}}\,{y_{{2}}}^{2}y_{{5}}-{\frac {7}{9}}\,{y_{{3}}
}^{2}y_{{6}}-{\frac {16}{9}}\,y_{{3}}y_{{7}}\nonumber
\end{eqnarray}
\begin{eqnarray}
&&-4/3\,y_{{5}}{y_{{1}}}^{2}
-4/3\,y_{{6}}{y_{{1}}}^{2}+1/3\,y_{{6}}{y_{{2}}}^{2}+1/3\,{y_{{3}}}^{2
}y_{{5}}- \left(  \left( 4/3\,y_{{3}}y_{{1}}{y_{{2}}}^{2}-8/3\,y_{{3}}
y_{{2}}{y_{{1}}}^{2}\right.\right.\nonumber\\
&&\left.\left.+4/3\,y_{{2}}y_{{1}}{y_{{3}}}^{2}-4\,y_{{1}}y_{{2}
}y_{{5}}-4\,y_{{1}}y_{{2}}y_{{4}}-4\,y_{{1}}y_{{3}}y_{{4}}-4/3\,y_{{1}
}y_{{6}}y_{{2}}-4\,y_{{1}}y_{{3}}y_{{6}}+8\,y_{{2}}y_{{3}}y_{{6}}
\right.\right.\nonumber\\
&&\left.\left.+8/3
\,y_{{2}}y_{{3}}y_{{4}}+8\,y_{{2}}y_{{3}}y_{{5}}-4/3\,y_{{1}}y_{{3}}y_
{{5}}-4/3\,y_{{5}}y_{{4}}-4/3\,y_{{6}}y_{{4}}+8/3\,y_{{6}}y_{{5}}-8/3
\,{y_{{1}}}^{2}{y_{{2}}}^{2}\right.\right.\nonumber\\
&&\left.\left.-8/3\,{y_{{1}}}^{2}{y_{{3}}}^{2}+16/3\,{y_
{{2}}}^{2}{y_{{3}}}^{2}-4\,{y_{{1}}}^{3}y_{{2}}-4\,{y_{{1}}}^{3}y_{{3}
}+4\,{y_{{3}}}^{3}y_{{2}}+4\,{y_{{2}}}^{3}y_{{3}}-8/3\,y_{{1}}y_{{7}}
\right.\right.\nonumber\\
&&\left.\left.+
4/3\,y_{{2}}y_{{8}}+4/3\,y_{{3}}y_{{9}}-8/3\,y_{{8}}y_{{1}}+4/3\,y_{{2
}}y_{{7}}-8/3\,y_{{9}}y_{{1}}+4/3\,y_{{9}}y_{{2}}+4/3\,y_{{3}}y_{{8}}-
8/3\,{y_{{4}}}^{2}\right.\right.\nonumber\\
&&\left.\left.+4/3\,{y_{{5}}}^{2}+4/3\,{y_{{6}}}^{2}+4/3\,{y_{{2}}
}^{4}+4/3\,{y_{{3}}}^{4}-{\frac {32}{3}}\,{y_{{1}}}^{2}y_{{4}}+16/3\,{
y_{{2}}}^{2}y_{{5}}+16/3\,{y_{{3}}}^{2}y_{{6}}\right.\right.\nonumber\\
&&\left.\left.+4/3\,y_{{3}}y_{{7}}-4\,
y_{{5}}{y_{{1}}}^{2}-4\,y_{{6}}{y_{{1}}}^{2}+4\,y_{{6}}{y_{{2}}}^{2}+4
\,{y_{{3}}}^{2}y_{{5}}-8/3\,{y_{{1}}}^{4} \right) \beta+\left(-1/3\,
{y_{{3}}}^{2}-1/3\,y_{{5}}\right.\right.\nonumber\\
&&\left.\left.-1/3\,{y_{{2}}}^{2}-2/3\,y_{{3}}y_{{2}}-1/3
\,y_{{6}}+2/3\,{y_{{1}}}^{2}+2/3\,y_{{4}}+1/3\,y_{{2}}y_{{1}}+1/3\,y_{
{3}}y_{{1}}\right)/G \right) {\alpha}^{-1}\nonumber\\ 
&&+{\frac {7}{18}}\,{y_{{1}}}^{4}\nonumber\\
\frac{d^4}{dt^4}w_{{2}}
&=&-{\frac {19}{9}}\,y_{{3}}y_{{1}}{y_{{2}}}^{2}+{\frac {11}{9}}\,y_{{3}}
y_{{2}}{y_{{1}}}^{2}+{\frac {11}{9}}\,y_{{2}}y_{{1}}{y_{{3}}}^{2}-13/3
\,y_{{1}}y_{{2}}y_{{5}}+5/3\,y_{{1}}y_{{2}}y_{{4}}+2/3\,y_{{1}}y_{{3}}
y_{{4}}\nonumber\\
&&+1/9\,y_{{1}}y_{{6}}y_{{2}}+2/3\,y_{{1}}y_{{3}}y_{{6}}+5/3\,y_{
{2}}y_{{3}}y_{{6}}+1/9\,y_{{2}}y_{{3}}y_{{4}}-13/3\,y_{{2}}y_{{3}}y_{{
5}}-{\frac {14}{9}}\,y_{{1}}y_{{3}}y_{{5}}\nonumber\\
&&-\left( {\frac {1}{108}}\,y
_{{3}}y_{{1}}{y_{{2}}}^{2}+{\frac {1}{108}}\,y_{{3}}y_{{2}}{y_{{1}}}^{
2}+{\frac {1}{108}}\,y_{{2}}y_{{1}}{y_{{3}}}^{2}+{\frac {1}{54}}\,y_{{
1}}y_{{6}}y_{{2}}+{\frac {1}{54}}\,y_{{2}}y_{{3}}y_{{4}}\right.\nonumber\\
&&\left.+{\frac {1}{54
}}\,y_{{1}}y_{{3}}y_{{5}}-{\frac {1}{108}}\,y_{{5}}y_{{4}}-{\frac {1}{
108}}\,y_{{6}}y_{{4}}-{\frac
  {1}{108}}\,y_{{6}}y_{{5}}+1/27\,{y_{{1}}}
^{2}{y_{{2}}}^{2}+1/27\,{y_{{1}}}^{2}{y_{{3}}}^{2}\right.\nonumber\\
&&\left.+1/27\,{y_{{2}}}^{2}
{y_{{3}}}^{2}-1/36\,{y_{{3}}}^{3}y_{{1}}-1/36\,{y_{{1}}}^{3}y_{{2}}-1/
36\,{y_{{1}}}^{3}y_{{3}}-1/36\,{y_{{2}}}^{3}y_{{1}}-1/36\,{y_{{3}}}^{3
}y_{{2}}\right.\nonumber\\
&&\left.-1/36\,{y_{{2}}}^{3}y_{{3}}-{\frac {1}{54}}\,y_{{1}}y_{{7}}-{
\frac {1}{54}}\,y_{{2}}y_{{8}}-{\frac {1}{54}}\,y_{{3}}y_{{9}}+{\frac 
{1}{108}}\,y_{{8}}y_{{1}}+{\frac {1}{108}}\,y_{{2}}y_{{7}}+{\frac {1}{
108}}\,y_{{9}}y_{{1}}\right.\nonumber\\
&&\left.+{\frac
  {1}{108}}\,y_{{9}}y_{{2}}+{\frac {1}{108}
}\,y_{{3}}y_{{8}}+{\frac {1}{108}}\,{y_{{4}}}^{2}+{\frac {1}{108}}\,{y
_{{5}}}^{2}+{\frac {1}{108}}\,{y_{{6}}}^{2}+{\frac {1}{108}}\,{y_{{2}}
}^{4}+{\frac {1}{108}}\,{y_{{3}}}^{4}\right.\nonumber\\
&&\left.-{\frac {1}{54}}\,{y_{{1}}}^{2}y_
{{4}}-{\frac {1}{54}}\,{y_{{2}}}^{2}y_{{5}}-{\frac {1}{54}}\,{y_{{3}}}
^{2}y_{{6}}+{\frac {1}{108}}\,y_{{3}}y_{{7}}+{\frac {1}{108}}\,{y_{{1}
}}^{4} \right) \alpha{\beta}^{-1}-\left(-1/18\,y_{{6}}\right.\nonumber\\
&&\left.-1/18\,y_{{5}}
+1/6\,\Lambda-1/18\,{y_{{3}}}^{2}-1/36\,y_{{3}}y_{{2}}-1/18\,{y_{{2}}
}^{2}-1/18\,y_{{4}}-1/36\,y_{{3}}y_{{1}}\right.\nonumber\\
&&\left.-1/18\,{y_{{1}}}^{2}-1/36\,y_{
{2}}y_{{1}}\right)/(G\beta)-{\frac {11}{9}}\,y_{{5}}y_{{4}}+1/9\,y_{{6}}y_{{
4}}-{\frac {11}{9}}\,y_{{6}}y_{{5}}-{\frac {5}{18}}\,{y_{{1}}}^{2}{y_{
{2}}}^{2}\nonumber\\
&&+{\frac {7}{18}}\,{y_{{1}}}^{2}{y_{{3}}}^{2}-{\frac {5}{18}}
\,{y_{{2}}}^{2}{y_{{3}}}^{2}+2/3\,{y_{{1}}}^{3}y_{{2}}-2/3\,{y_{{2}}}^
{3}y_{{1}}+2/3\,{y_{{3}}}^{3}y_{{2}}-2/3\,{y_{{2}}}^{3}y_{{3}}\nonumber\\
&&-1/9\,y_
{{1}}y_{{7}}-{\frac {22}{9}}\,y_{{2}}y_{{8}}-1/9\,y_{{3}}y_{{9}}-{
\frac {16}{9}}\,y_{{8}}y_{{1}}-1/9\,y_{{2}}y_{{7}}+2/9\,y_{{9}}y_{{1}}
-1/9\,y_{{9}}y_{{2}}\nonumber\\
&&-{\frac {16}{9}}\,y_{{3}}y_{{8}}-{\frac {5}{18}}\,
{y_{{4}}}^{2}-{\frac {29}{18}}\,{y_{{5}}}^{2}-{\frac {5}{18}}\,{y_{{6}
}}^{2}+{\frac {7}{18}}\,{y_{{2}}}^{4}-{\frac {5}{18}}\,{y_{{3}}}^{4}-{
\frac {7}{9}}\,{y_{{1}}}^{2}y_{{4}}-1/9\,{y_{{2}}}^{2}y_{{5}}\nonumber
\end{eqnarray}
\begin{eqnarray}
&&-{\frac {
7}{9}}\,{y_{{3}}}^{2}y_{{6}}+2/9\,y_{{3}}y_{{7}}+1/3\,y_{{6}}{y_{{1}}}
^{2}-4/3\,y_{{6}}{y_{{2}}}^{2}-{\frac {5}{18}}\,{y_{{1}}}^{4}-4/3\,{y_
{{2}}}^{2}y_{{4}}+1/3\,{y_{{3}}}^{2}y_{{4}}\nonumber\\
&&- \left(  \left( -8/3\,y_{{
3}}y_{{1}}{y_{{2}}}^{2}+4/3\,y_{{3}}y_{{2}}{y_{{1}}}^{2}+4/3\,y_{{2}}y
_{{1}}{y_{{3}}}^{2}-4\,y_{{1}}y_{{2}}y_{{5}}-4\,y_{{1}}y_{{2}}y_{{4}}+
8\,y_{{1}}y_{{3}}y_{{4}}\right.\right.\nonumber\\
&&\left.\left.-4/3\,y_{{1}}y_{{6}}y_{{2}}+8\,y_{{1}}y_{{3}}y
_{{6}}-4\,y_{{2}}y_{{3}}y_{{6}}-4/3\,y_{{2}}y_{{3}}y_{{4}}-4\,y_{{2}}y
_{{3}}y_{{5}}+8/3\,y_{{1}}y_{{3}}y_{{5}}-4/3\,y_{{5}}y_{{4}}\right.\right.\nonumber\\
&&\left.\left.+8/3\,y_{{
6}}y_{{4}}-4/3\,y_{{6}}y_{{5}}-8/3\,{y_{{1}}}^{2}{y_{{2}}}^{2}+16/3\,{
y_{{1}}}^{2}{y_{{3}}}^{2}-8/3\,{y_{{2}}}^{2}{y_{{3}}}^{2}+4\,{y_{{3}}}
^{3}y_{{1}}\right.\right.\nonumber\\
&&\left.\left.+4\,{y_{{1}}}^{3}y_{{3}}-4\,{y_{{2}}}^{3}y_{{1}}-4\,{y_{{2}
}}^{3}y_{{3}}+4/3\,y_{{1}}y_{{7}}-8/3\,y_{{2}}y_{{8}}+4/3\,y_{{3}}y_{{
9}}+4/3\,y_{{8}}y_{{1}}-8/3\,y_{{2}}y_{{7}}\right.\right.\nonumber\\
&&\left.\left.+4/3\,y_{{9}}y_{{1}}-8/3\,y
_{{9}}y_{{2}}+4/3\,y_{{3}}y_{{8}}+4/3\,{y_{{4}}}^{2}-8/3\,{y_{{5}}}^{2
}+4/3\,{y_{{6}}}^{2}-8/3\,{y_{{2}}}^{4}+4/3\,{y_{{3}}}^{4}\right.\right.\nonumber\\
&&\left.\left.+16/3\,{y_{{
1}}}^{2}y_{{4}}-{\frac {32}{3}}\,{y_{{2}}}^{2}y_{{5}}+16/3\,{y_{{3}}}^
{2}y_{{6}}+4/3\,y_{{3}}y_{{7}}+4\,y_{{6}}{y_{{1}}}^{2}-4\,y_{{6}}{y_{{
2}}}^{2}+4/3\,{y_{{1}}}^{4}\right.\right.\nonumber\\
&&\left.\left.-4\,{y_{{2}}}^{2}y_{{4}}+4\,{y_{{3}}}^{2}y_
{{4}} \right) \beta+\left(-1/3\,{y_{{3}}}^{2}+2/3\,y_{{5}}+2/3\,{y_{
{2}}}^{2}+1/3\,y_{{3}}y_{{2}}-1/3\,y_{{6}}\right.\right.\nonumber\\
&&\left.\left.-1/3\,{y_{{1}}}^{2}-1/3\,y_{
{4}}+1/3\,y_{{2}}y_{{1}}-2/3\,y_{{3}}y_{{1}}\right)/G \right) {\alpha}^{-1
}\nonumber\\
%\end{eqnarray}
%\begin{eqnarray}
\frac{d^4}{dt^4}w_{{3}}&=&
{\frac {11}{9}}\,y_{{3}}y_{{1}}{y_{{2}}}^{2}+{\frac {11}{9}}\,y_{{3}}y
_{{2}}{y_{{1}}}^{2}-{\frac {19}{9}}\,y_{{2}}y_{{1}}{y_{{3}}}^{2}+2/3\,
y_{{1}}y_{{2}}y_{{5}}+2/3\,y_{{1}}y_{{2}}y_{{4}}\nonumber\\
&&+5/3\,y_{{1}}y_{{3}}y_
{{4}}-{\frac {14}{9}}\,y_{{1}}y_{{6}}y_{{2}}-13/3\,y_{{1}}y_{{3}}y_{{6
}}-13/3\,y_{{2}}y_{{3}}y_{{6}}+1/9\,y_{{2}}y_{{3}}y_{{4}}+5/3\,y_{{2}}
y_{{3}}y_{{5}}\nonumber\\
&&+1/9\,y_{{1}}y_{{3}}y_{{5}}- \left( {\frac {1}{108}}\,y_
{{3}}y_{{1}}{y_{{2}}}^{2}+{\frac {1}{108}}\,y_{{3}}y_{{2}}{y_{{1}}}^{2
}+{\frac {1}{108}}\,y_{{2}}y_{{1}}{y_{{3}}}^{2}+{\frac {1}{54}}\,y_{{1
}}y_{{6}}y_{{2}}\right.\nonumber\\
&&\left.+{\frac {1}{54}}\,y_{{2}}y_{{3}}y_{{4}}+{\frac {1}{54}
}\,y_{{1}}y_{{3}}y_{{5}}-{\frac {1}{108}}\,y_{{5}}y_{{4}}-{\frac {1}{
108}}\,y_{{6}}y_{{4}}-{\frac {1}{108}}\,y_{{6}}y_{{5}}+1/27\,{y_{{1}}}
^{2}{y_{{2}}}^{2}\right.\nonumber\\
&&\left.+1/27\,{y_{{1}}}^{2}{y_{{3}}}^{2}+1/27\,{y_{{2}}}^{2}
{y_{{3}}}^{2}-1/36\,{y_{{3}}}^{3}y_{{1}}-1/36\,{y_{{1}}}^{3}y_{{2}}-1/
36\,{y_{{1}}}^{3}y_{{3}}-1/36\,{y_{{2}}}^{3}y_{{1}}\right.\nonumber\\
&&\left.-1/36\,{y_{{3}}}^{3
}y_{{2}}-1/36\,{y_{{2}}}^{3}y_{{3}}-{\frac {1}{54}}\,y_{{1}}y_{{7}}-{
\frac {1}{54}}\,y_{{2}}y_{{8}}-{\frac {1}{54}}\,y_{{3}}y_{{9}}+{\frac 
{1}{108}}\,y_{{8}}y_{{1}}+{\frac {1}{108}}\,y_{{2}}y_{{7}}\right.\nonumber\\
&&\left.+{\frac {1}{
108}}\,y_{{9}}y_{{1}}+{\frac {1}{108}}\,y_{{9}}y_{{2}}+{\frac {1}{108}
}\,y_{{3}}y_{{8}}+{\frac {1}{108}}\,{y_{{4}}}^{2}+{\frac {1}{108}}\,{y
_{{5}}}^{2}+{\frac {1}{108}}\,{y_{{6}}}^{2}+{\frac {1}{108}}\,{y_{{2}}
}^{4}\right.\nonumber\\
&&\left.+{\frac {1}{108}}\,{y_{{3}}}^{4}-{\frac {1}{54}}\,{y_{{1}}}^{2}y_
{{4}}-{\frac {1}{54}}\,{y_{{2}}}^{2}y_{{5}}-{\frac {1}{54}}\,{y_{{3}}}
^{2}y_{{6}}+{\frac {1}{108}}\,y_{{3}}y_{{7}}+{\frac {1}{108}}\,{y_{{1}
}}^{4} \right) \alpha{\beta}^{-1}-\left(-1/18\,y_{{6}}\right.\nonumber\\
&&\left.-1/18\,y_{{5}}
+1/6\,\Lambda-1/18\,{y_{{3}}}^{2}-1/36\,y_{{3}}y_{{2}}-1/18\,{y_{{2}}
}^{2}-1/18\,y_{{4}}-1/36\,y_{{3}}y_{{1}}\right.\nonumber\\
&&\left.-1/18\,{y_{{1}}}^{2}-1/36\,y_{
{2}}y_{{1}}\right)/(G\beta)+1/9\,y_{{5}}y_{{4}}-{\frac {11}{9}}\,y_{{6}}y_{{
4}}-{\frac {11}{9}}\,y_{{6}}y_{{5}}+{\frac {7}{18}}\,{y_{{1}}}^{2}{y_{
{2}}}^{2}-{\frac {5}{18}}\,{y_{{1}}}^{2}{y_{{3}}}^{2}\nonumber\\
&&-{\frac {5}{18}}
\,{y_{{2}}}^{2}{y_{{3}}}^{2}-2/3\,{y_{{3}}}^{3}y_{{1}}+2/3\,{y_{{1}}}^
{3}y_{{3}}-2/3\,{y_{{3}}}^{3}y_{{2}}+2/3\,{y_{{2}}}^{3}y_{{3}}-1/9\,y_
{{1}}y_{{7}}-1/9\,y_{{2}}y_{{8}}\nonumber\\
&&-{\frac {22}{9}}\,y_{{3}}y_{{9}}+2/9\,
y_{{8}}y_{{1}}+2/9\,y_{{2}}y_{{7}}-{\frac {16}{9}}\,y_{{9}}y_{{1}}-{
\frac {16}{9}}\,y_{{9}}y_{{2}}-1/9\,y_{{3}}y_{{8}}-{\frac {5}{18}}\,{y
_{{4}}}^{2}-{\frac {5}{18}}\,{y_{{5}}}^{2}\nonumber\\
&&-{\frac {29}{18}}\,{y_{{6}}}
^{2}-{\frac {5}{18}}\,{y_{{2}}}^{4}+{\frac {7}{18}}\,{y_{{3}}}^{4}-{
\frac {7}{9}}\,{y_{{1}}}^{2}y_{{4}}-{\frac {7}{9}}\,{y_{{2}}}^{2}y_{{5
}}-1/9\,{y_{{3}}}^{2}y_{{6}}-1/9\,y_{{3}}y_{{7}}+1/3\,y_{{5}}{y_{{1}}}
^{2}\nonumber\\
&&-4/3\,{y_{{3}}}^{2}y_{{5}}- \left(  \left( 4/3\,y_{{3}}y_{{1}}{y_{
{2}}}^{2}+4/3\,y_{{3}}y_{{2}}{y_{{1}}}^{2}-8/3\,y_{{2}}y_{{1}}{y_{{3}}
}^{2}+8\,y_{{1}}y_{{2}}y_{{5}}+8\,y_{{1}}y_{{2}}y_{{4}}
\right.\right.\nonumber
\end{eqnarray}
\begin{eqnarray}
&&\left.\left.-4\,y_{{1}}y_{{
3}}y_{{4}}+8/3\,y_{{1}}y_{{6}}y_{{2}}-4\,y_{{1}}y_{{3}}y_{{6}}-4\,y_{{
2}}y_{{3}}y_{{6}}-4/3\,y_{{2}}y_{{3}}y_{{4}}-4\,y_{{2}}y_{{3}}y_{{5}}-
4/3\,y_{{1}}y_{{3}}y_{{5}}\right.\right.\nonumber\\
&&\left.\left.+8/3\,y_{{5}}y_{{4}}-4/3\,y_{{6}}y_{{4}}-4/3
\,y_{{6}}y_{{5}}+16/3\,{y_{{1}}}^{2}{y_{{2}}}^{2}-8/3\,{y_{{1}}}^{2}{y
_{{3}}}^{2}-8/3\,{y_{{2}}}^{2}{y_{{3}}}^{2}\right.\right.\nonumber\\
&&\left.\left.-4\,{y_{{3}}}^{3}y_{{1}}+4
\,{y_{{1}}}^{3}y_{{2}}+4\,{y_{{2}}}^{3}y_{{1}}-4\,{y_{{3}}}^{3}y_{{2}}
+4/3\,y_{{1}}y_{{7}}+4/3\,y_{{2}}y_{{8}}-8/3\,y_{{3}}y_{{9}}+4/3\,y_{{
8}}y_{{1}}\right.\right.\nonumber\\
&&\left.\left.+4/3\,y_{{2}}y_{{7}}+4/3\,y_{{9}}y_{{1}}+4/3\,y_{{9}}y_{{2}}
-8/3\,y_{{3}}y_{{8}}+4/3\,{y_{{4}}}^{2}+4/3\,{y_{{5}}}^{2}-8/3\,{y_{{6
}}}^{2}+4/3\,{y_{{2}}}^{4}\right.\right.\nonumber\\
&&\left.\left.-8/3\,{y_{{3}}}^{4}+16/3\,{y_{{1}}}^{2}y_{{4
}}+16/3\,{y_{{2}}}^{2}y_{{5}}-{\frac {32}{3}}\,{y_{{3}}}^{2}y_{{6}}-8/
3\,y_{{3}}y_{{7}}+4\,y_{{5}}{y_{{1}}}^{2}-4\,{y_{{3}}}^{2}y_{{5}}
\right.\right.\nonumber\\
&&\left.\left.+4/3
\,{y_{{1}}}^{4}+4\,{y_{{2}}}^{2}y_{{4}}-4\,{y_{{3}}}^{2}y_{{4}}
 \right) \beta+\left(2/3\,{y_{{3}}}^{2}-1/3\,y_{{5}}-1/3\,{y_{{2}}}^
{2}+1/3\,y_{{3}}y_{{2}}+2/3\,y_{{6}}\right.\right.\nonumber\\
&&\left.\left.-1/3\,{y_{{1}}}^{2}-1/3\,y_{{4}}-2
/3\,y_{{2}}y_{{1}}+1/3\,y_{{3}}y_{{1}}\right)/G \right) {\alpha}^{-1}-{
\frac{5}{18}}\,{y_{{1}}}^{4}+1/3\,{y_{{2}}}^{2}y_{{4}}\nonumber\\
&&-4/3\,{y_{{3}}}^{2}y_{{4}}\label{A10}
\end{eqnarray}

\section{Linearized tensorial and scalar fields \label{ap3}}
The first order linearized Ricci and Einstein tensor read 
\begin{eqnarray*}
&&R_{ab}^1=\frac{1}{2}\left\{h^n_{a|bn}+h^n_{b|an}-\square h_{ab}-h^n_{n|ab}\right\}\\
&&G_{ab}^1=\frac{1}{2}\left\{h^n_{a|bn}+h^n_{b|an}-\square h_{ab}-\eta_{ab}\left(h^{mn}_{|mn}-\square h^n_n \right)\right\}.
\end{eqnarray*}

Also following (\ref{ecl}) we have 
\begin{eqnarray*}
&&H^1_{ab}=2\left\{h^{nm}_{|mnab}-\square h^n_{n|ab}-\eta_{ab}\left(\square h^{mn}_{mn} -\square^2h^n_n\right)\right\}\\
&&H^2_{ab}=\frac{1}{2}H^1_{ab}-\square G^1_{ab}
\end{eqnarray*}

The Einstein tensor can now be thought of as an operator acting on an arbitrary tensor field $\theta_{rs}$ 
\[
G^1_{ab}(\theta_{rs})=\frac{1}{2}\left\{\theta^n_{a|bn}+\theta^n_{b|an}-\square \theta_{ab}-\eta_{ab}\left(\theta^{mn}_{|mn}-\square \theta^n_n \right)\right\}.
\]
With this in mind, it is a simple algebraic exercise just to evaluate that
\begin{eqnarray*}
&&G^1_{ab}(\eta_{rs}R^1)=-\frac{1}{2}H^1_{ab}\\
&&G^1_{ab}(R^1_{rs})=-\frac{1}{2}\square G^1_{ab}(h_{rs})
\end{eqnarray*}
where $R^1_{ab}$ is the Ricci tensor and $R^1$ is the Riemann scalar. Now the prescription of oscillator variables adopted by Stelle \cite{Stelle1978} can be followed, so that in first order perturbation we have
\[
\mathcal{L}=\frac{1}{2}\left\{h^{ab}G^1_ab(\Sigma_{rs})-\frac{1}{2}\Sigma_{ab}\Sigma^{ab}+a\Sigma^{ab}h_{ab}-\frac{a^2}{2}h_{ab}h^{ab}+q\Sigma^n_n\Sigma^m_m-2aqh^n_n\Sigma^m_m+a^2qh^n_nh^m_m\right\}.
\]
Variations with respect to the fields $\Sigma_{ab}$ yield 
 \[
 \Sigma_{ab}=G^1_{ab}(h_{rs})+ah_{ab}+\frac{2q}{8q-1}R^1\eta_{ab},
 \]
 and variations with respect to $h_{ab}$ yield 
 \[
 \frac{H^2_{ab}}{2}+2aG^1_{ab}(h_{ab})-\frac{q}{8q-1}H^1_{ab}=0.
 \]
 Comparing with the field equations (\ref{tensor E1}) we obtain the values for $a$ and $q$
 \begin{eqnarray*}
 &&a=\frac{1}{4\alpha}\\
 &&q=\frac{\alpha/3-\beta}{2\alpha/3-8\beta}.
 \end{eqnarray*}
 The tensor fields $\phi$ and $\psi$ are defined in the following $\Sigma_{ab}=\phi_{ab}+\psi_{ab}$ and $h_{ab}=(\phi_{ab}-\psi_{ab})/a$ so that the following action is obtained 
 \begin{eqnarray}
 \mathcal{L}=\frac{1}{2}\left\{\phi^{ab}G^1_{ab}(\phi_{rs})-\psi^{ab}G^1_{ab}(\psi_{rs})+\frac{m^2_2}{2}(\psi_{ab}\psi^{ab}-2q(\psi^n_n)^2)\right\}, \label{lag.}
 \end{eqnarray}
 where 
 \begin{equation}
 m_2=\frac{1}{\sqrt{\alpha}} \label{m2}
 \end{equation}
 is the mass of the spin 2 field as we will show in the following. Compared to the Fierz-Pauli lagrangian, there is an overall sign difference due to our sign convention on the Riemann tensor. We quote the correct Fierz-Pauli lagrangian for a massive spin 2 field of mass $m$, for example $\xi_{ab}$
\[
\mathcal{L}^{FP}=\frac{1}{2}\left\{ \frac{1}{2}\xi^{ab}\square \xi_{ab}-\frac{1}{2}\xi^m_m\square \xi^n_n+\xi^{ab}\xi^n_{n|ab}-\xi^{ab}\xi^n_{a|bn}-\frac{m^2}{2}\left(\xi_{ab}\xi^{ab}-(\xi^n_n)^2\right)\right\},  
\]
and as it is well known, if the numerical factors do not match exactly the above expression there is a scalar field, $\chi$ with negative energy $T^\chi_{00}$ \footnote{to see the details on how to extract the scalar component of (\ref{lag.}), see the appendix of van Dam and Veltman's article \cite{Veltman}.}. The first part of (\ref{lag.}) corresponds to the mass less spin 2 field $\phi_{ab}$ with positive energy density $T^{\phi}_{00}>0$. In this case the entire $\psi_{ab}$ field has the wrong sign in (\ref{lag.}), being a ghost $T^\psi_{00}<0$, so that the scalar field in this case will come out with positive energy. To see this we must take the trace and divergences of the filed equation obtained by varying (\ref{lag.}) with respect to $\psi_{ab}$ 
\[
\square \psi^n_n-\frac{m^2_2(8q-1)}{2(1-2q)}\psi^n_n=0 \rightarrow m_0=\frac{m^2_2(8q-1)}{2(1-2q)},
\]
where 
\begin{eqnarray}
m_0=\sqrt{\frac{1}{-6\beta}} \label{m0}
\end{eqnarray}
is the mass of the scalar field. Substituting the above expression back into the equation of motion yields for the trace less part of $\psi_{ab}$, $\psi^t_{ab}=\psi_{ab}-\eta_{ab}\psi^n_n/4$
\[
\square \psi^t_{ab}-m_2^2\psi^t_{ab}-(4q-1)\left(\psi^n_{n|ab} -\frac{1}{4}m_0^2\psi^n_n\eta_{ab}\right)=0.
\] 
Since the trace of the above equation is zero by construction, considering a traceless $\theta_{ab}$ it can be inferred that $\psi^t_{ab}$
\[
\psi^t_{ab}=\theta_{ab}+\frac{2(1-2q)}{3m^2_2}\left(\psi^n_{n|ab} -\frac{m_0^2}{4}\psi^n_n\eta_{ab}\right).
\]
Substituting the above expression in the previous one, it can be checked explicitly that $\square \theta_{ab}-m_2^2\theta_{ab}=0$, $\theta^a_a=0$ and $\theta^n_{a|n}=0$, being the pure spin 2 massive field. The $\psi_{ab}$ field in (\ref{lag.}) can then be written as
\begin{eqnarray}
&&\psi_{ab}=\theta_{ab}+\frac{(1-2q)}{3}\left[ \eta_{ab}\psi^n_n+\frac{2}{m_2^2}\psi^n_{n|ab}\right]\nonumber\\
&&\psi^n_n=\frac{3}{8q-1}\theta^n_n+\frac{1}{m_0^2}\square\psi^n_n,\label{p.escalar}
\end{eqnarray}
where $m_0$ is the mass of the scalar field. The expression for $\psi_{ab}$ and its trace, (\ref{p.escalar}) can be substituted back into (\ref{lag.}) eliminating any coupling between $\theta_{ab}$ and $\psi^n_n$ and the higher derivative terms. Which results without the use of the equation of motion, in the following lagrangian 
\[
\mathcal{L}=\frac{1}{2}\left\{\phi^{ab}G^1_{ab}(\phi_{rs})-\theta^{ab}G^1_{ab}(\theta_{rs})+\frac{m_2^2}{2}(\theta_{ab}\theta^{ab}-(\theta^n_n)^2)+\frac{1}{3}(1-2q)^2\left[\psi^n_n\square \psi^n_n-m_0^2(\psi^n_n)^2 \right]\right\}.
\]

As usual, in the linearized approach, the coupling to matter or other fields is through the energy momentum tensor 
\[
\kappa \left( \frac{\phi_{ab}-\psi_{ab}}{a}\right)T^{ab}=\kappa 4\alpha \left[ \phi_{ab}-\theta_{ab}+\left(\frac{\beta}{\alpha/3-4\beta}\right)\psi_n^n\eta_{ab}\right]T^{ab},
\]
where the gradient term was absorbed into a gauge transformation of the $\phi_{ab}$ field and the scalar part couples to the trace of the energy momentum tensor. In this  present work only vacuum solutions are considered so that as long as the amplitude of the perturbation $h_{ab}$ does not increase much, the fields involved behave as entirely free, non coupled. Written with respect to the old field variables, we have 
\begin{eqnarray}
&&\phi_{ab}=\frac{1}{4\alpha}h_{ab}+\frac{1}{2}R^1_{ab}-\left(\frac{\alpha/3+2\beta}{4\alpha}\right)R^1\eta_{ab}\nonumber\\
&&\theta_{ab}=\frac{1}{2}R^1_{ab}-\frac{1}{12}R^1\eta_{ab}+\frac{\beta}{\alpha m_2^2}R^1_{|ab}\nonumber\\
&&\psi^n_n=\left(\frac{\alpha-12\beta}{6\alpha}\right)R^1\label{campos220}
\end{eqnarray}
\section{Linearized system \label{ap2}}
Here we obtain the linearized differential equations according to section \ref{sec.perturb}. In table \ref{tabela}, we show the numerical initial condition, for the complete non perturbed solution shown in Figure \ref{fig1_1}, for the time instant $t=8.000063354e7$. 

\begin{table}[h]
\begin{tabular}{|c|c|c|c|c|c|}
\hline
$a_1$ &$ a_2$ & $a_3$  \\
2.752145223e4 & 9.351773627e4 & 6.987130075e4 \\
\hline
$\dot{a}_1$ & $\dot{a}_2$ & $\dot{a}_3$\\
9.965181851e-5 & 2.525829890e-4 & 
1.942842750e-4\\
\hline
$\ddot{a}_1$ & $\ddot{a}_2$ & $\ddot{a}_3$ \\ 
1.017372156e-4 & 2.153460508e-4 & 1.352492057 e-4\\
\hline
$\dddot{a}_1$ & $\dddot{a}_2$ & $\dddot{a}_3$\\ 
1.623256323e-5 & 9.817489598e-5 & 
7.056676908e-5\\
\hline
$c_1$ & $c_2$ & $c_3$ \\
9.173663e3 & 3.117199e4 & 2.328999e4\\
\hline
 $b_1$ & $b_2$ & $b_3$\\
 2.293455e-4 & 7.793157e-4 & 5.822617e-4\\
\hline
$\alpha_1$ & $\alpha_2$ & $\alpha_3$ \\
0.148260 & 0.503708 & 0.376357 \\
\hline
$\gamma_1$ & $\gamma_2$ & $\gamma_3$ \\
0.666656 & 0.6666648 & 0.6666627\\
\hline
\end{tabular}
\caption{Initial condition and numerical values for the complete non perturbed solution shown in Figure \ref{fig1_1} for the instant $t=8.000063354e7$. We also show the result of the linear regression, and the approximate, mean solution shown in Figure \ref{fig1_1}.\label{tabela}}
\end{table}

Remind that the perturbation is linearly related to the metric
while the scale factors are quadratic, then

\begin{eqnarray*}
&&\hat{h}_{1}=  a_{1}^{2}-(c_{1}+b_{1}t)^{2}\\
&&\hat{h}_{2}=  a_{2}^{2}-(c_{2}+b_{2}t)^{2}\\
&&\hat{h}_{3}=  a_{3}^{2}-(c_{3}+b_{3}t)^{2}\\
&&\dot{\hat{h}}_{1}=  2a_{1}\dot{a}_{1}-2b_{1}(c_{1}+b_{1}t)\\
&&\dot{\hat{h}}_{2}=  2a_{2}\dot{a}_{2}-2b_{2}(c_{2}+b_{2}t)\\
&&\dot{\hat{h}}_{3}=  2a_{3}\dot{a}_{3}-2b_{3}(c_{3}+b_{3}t)\\
&&\ddot{\hat{h}}_{1} = 2a_{1}\ddot{a}_{1}+2\dot{(a}_{1})^{2}-2b_{1}^{2}\\
&&\ddot{\hat{h}}_{2}=  2a_{2}\ddot{a}_{2}+2\dot{(a}_{2})^{2}-2b_{2}^{2}\\
&&\ddot{\hat{h}}_{3}=  2a_{3}\ddot{a}_{3}+2\dot{(a}_{3})^{2}-2b_{3}^{2}\\
&&\dddot{\hat{h}}_{1}=  2a_{1}\dddot{a}_{1}+6\dot{a}_{1}\ddot{a}_{1}\\
%\end{eqnarray*}
%\begin{eqnarray*}
&&\dddot{\hat{h}}_{2}=  2a_{2}\dddot{a}_{2}+6\dot{a}_{2}\ddot{a}_{2}\\
&&\dddot{\hat{h}}_{3}=  2a_{3}\dddot{a}_{3}+6\dot{a}_{3}\ddot{a}_{3}\\
&&\frac{d^{4}}{dt^{4}}\hat{h}_{1}=  2a_{1}\frac{d^{4}}{dt^{4}}a_{1}+8\dot{a}_{1}\dddot{a}_{1}+6(\dot{a}_{1})^{2}\\
&&\frac{d^{4}}{dt^{4}}h_{2}=  2a_{2}\frac{d^{4}}{dt^{4}}a_{2}+8\dot{a}_{2}\dddot{a}_{2}+6(\dot{a}_{2})^{2}\\
&&\frac{d^{4}}{dt^{4}}h_{3}=
2a_{3}\frac{d^{4}}{dt^{4}}a_{3}+8\dot{a}_{3}\dddot{a}_{3}+6(\dot{a}_{3})^{2}.
\end{eqnarray*}
Now the differential equations given in eq. (\ref{equacao de movimento linear}) read
\begin{eqnarray*}
 &&y_{1} = \hat{h}_{1}\\
 &&y_{2} = \dot{\hat{h}}_{1}\\
 &&y_{3} = \ddot{\hat{h}}_{1}\\
 &&y_{4} = \dddot{\hat{h}}_{1}\\
 &&y_{5} =\hat{h}_{2}\\
 &&y_{6} = \dot{\hat{h}}_{2}\\
 &&y_{7} = \ddot{\hat{h}}_{2}\\
 &&y_{8} = \dddot{\hat{h}}_{2}\\
 &&y_{9} = \hat{h}_{3}\\
 &&y_{10}=  \dot{\hat{h}}_{3}\\
 &&y_{11}=  \ddot{\hat{h}}_{3}\\
 &&y_{12}=  \dddot{\hat{h}}_{3}
 \end{eqnarray*}
 \begin{eqnarray*}
&&\frac{d^{4}}{dt^{4}}\hat{h}_{1}=  \left(12\,\beta\,\Lambda\,{c_{{1}}}^{2}{c_{{2}}}^{2}y_{{9}}-\alpha\,\Lambda\,{c_{{1}}}^{2}{c_{{2}}}^{2}y_{{9}}-\alpha\,\Lambda\,{c_{{2}}}^{2}{c_{{3}}}^{2}y_{{1}}+12\,\Lambda\,{c_{{1}}}^{2}{c_{{3}}}^{2}y_{{5}}\beta-\Lambda\,{c_{{1}}}^{2}{c_{{3}}}^{2}y_{{5}}\alpha-\right.\\
 && 24\,\beta\,\Lambda\,{c_{{2}}}^{2}{c_{{3}}}^{2}y_{{1}}-2\,\alpha\,\Lambda\,{c_{{2}}}^{2}{c_{{3}}}^{2}b_{{1}}c_{{1}}t-2\,\alpha\,\Lambda\,{c_{{1}}}^{2}{c_{{2}}}^{2}b_{{3}}c_{{3}}t+24\,\beta\,\Lambda\,{c_{{1}}}^{2}{c_{{2}}}^{2}b_{{3}}c_{{3}}t-\\
 && 48\,\beta\,\Lambda\,{c_{{2}}}^{2}{c_{{3}}}^{2}b_{{1}}c_{{1}}t+24\,\Lambda\,{c_{{1}}}^{2}{c_{{3}}}^{2}b_{{2}}c_{{2}}t\beta-2\,\Lambda\,{c_{{1}}}^{2}{c_{{3}}}^{2}b_{{2}}c_{{2}}t\alpha+{c_{{1}}}^{2}{c_{{2}}}^{2}y_{{11}}\alpha+\\
&&  6\,\beta\,{c_{{1}}}^{2}{c_{{3}}}^{2}y_{{7}}+\alpha\,{c_{{1}}}^{2}{c_{{3}}}^{2}y_{{7}}-12\,{c_{{2}}}^{2}{c_{{3}}}^{2}y_{{3}}\beta+12\,{c_{{1}}}^{2}{c_{{2}}}^{2}{b_{{3}}}^{2}\beta+2\,{c_{{2}}}^{2}{c_{{3}}}^{2}{b_{{1}}}^{2}\alpha-\\
 && 24\,{c_{{2}}}^{2}{c_{{3}}}^{2}{b_{{1}}}^{2}\beta+{c_{{2}}}^{2}{c_{{3}}}^{2}y_{{3}}\alpha+2\,\alpha\,{c_{{1}}}^{2}{c_{{3}}}^{2}{b_{{2}}}^{2}+2\,{c_{{1}}}^{2}{c_{{2}}}^{2}{b_{{3}}}^{2}\alpha+12\,\beta\,{c_{{1}}}^{2}{c_{{3}}}^{2}{b_{{2}}}^{2}+\\
&&  6\,{c_{{1}}}^{2}{c_{{2}}}^{2}y_{{11}}\beta-\alpha\,\Lambda\,{c_{{1}}}^{2}{c_{{2}}}^{2}{b_{{3}}}^{2}{t}^{2}-\alpha\,\Lambda\,{c_{{2}}}^{2}{c_{{3}}}^{2}{b_{{1}}}^{2}{t}^{2}+12\,\Lambda\,{c_{{1}}}^{2}{c_{{3}}}^{2}{b_{{2}}}^{2}{t}^{2}\beta-\\
 && \left.24\,\beta\,\Lambda\,{c_{{2}}}^{2}{c_{{3}}}^{2}{b_{{1}}}^{2}{t}^{2}+12\,\beta\,\Lambda\,{c_{{1}}}^{2}{c_{{2}}}^{2}{b_{{3}}}^{2}{t}^{2}-\Lambda\,{c_{{1}}}^{2}{c_{{3}}}^{2}{b_{{2}}}^{2}{t}^{2}\alpha\right)/\left(18\, G{c_{{3}}}^{2}{c_{{2}}}^{2}\alpha\,\beta\right)
 \end{eqnarray*}
 \begin{eqnarray*}
&&\frac{d^{4}}{dt^{4}}\hat{h}_{2}=  \left(12\,\beta\,\Lambda\,{c_{{1}}}^{2}{c_{{2}}}^{2}y_{{9}}-\alpha\,\Lambda\,{c_{{1}}}^{2}{c_{{2}}}^{2}y_{{9}}-\alpha\,\Lambda\,{c_{{2}}}^{2}{c_{{3}}}^{2}y_{{1}}-24\,\Lambda\,{c_{{1}}}^{2}{c_{{3}}}^{2}y_{{5}}\beta-\Lambda\,{c_{{1}}}^{2}{c_{{3}}}^{2}y_{{5}}\alpha+\right.\\
&&  12\,\beta\,\Lambda\,{c_{{2}}}^{2}{c_{{3}}}^{2}y_{{1}}-2\,\alpha\,\Lambda\,{c_{{2}}}^{2}{c_{{3}}}^{2}b_{{1}}c_{{1}}t-2\,\alpha\,\Lambda\,{c_{{1}}}^{2}{c_{{2}}}^{2}b_{{3}}c_{{3}}t+24\,\beta\,\Lambda\,{c_{{1}}}^{2}{c_{{2}}}^{2}b_{{3}}c_{{3}}t+\\
 && 24\,\beta\,\Lambda\,{c_{{2}}}^{2}{c_{{3}}}^{2}b_{{1}}c_{{1}}t-48\,\Lambda\,{c_{{1}}}^{2}{c_{{3}}}^{2}b_{{2}}c_{{2}}t\beta-2\,\Lambda\,{c_{{1}}}^{2}{c_{{3}}}^{2}b_{{2}}c_{{2}}t\alpha+{c_{{1}}}^{2}{c_{{2}}}^{2}y_{{11}}\alpha-\\
 && 12\,\beta\,{c_{{1}}}^{2}{c_{{3}}}^{2}y_{{7}}+\alpha\,{c_{{1}}}^{2}{c_{{3}}}^{2}y_{{7}}+6\,{c_{{2}}}^{2}{c_{{3}}}^{2}y_{{3}}\beta+12\,{c_{{1}}}^{2}{c_{{2}}}^{2}{b_{{3}}}^{2}\beta+2\,{c_{{2}}}^{2}{c_{{3}}}^{2}{b_{{1}}}^{2}\alpha+\\
 && 12\,{c_{{2}}}^{2}{c_{{3}}}^{2}{b_{{1}}}^{2}\beta+{c_{{2}}}^{2}{c_{{3}}}^{2}y_{{3}}\alpha+2\,\alpha\,{c_{{1}}}^{2}{c_{{3}}}^{2}{b_{{2}}}^{2}+2\,{c_{{1}}}^{2}{c_{{2}}}^{2}{b_{{3}}}^{2}\alpha-\\
 && 24\,\beta\,{c_{{1}}}^{2}{c_{{3}}}^{2}{b_{{2}}}^{2}+6\,{c_{{1}}}^{2}{c_{{2}}}^{2}y_{{11}}\beta-\alpha\,\Lambda\,{c_{{1}}}^{2}{c_{{2}}}^{2}{b_{{3}}}^{2}{t}^{2}-\alpha\,\Lambda\,{c_{{2}}}^{2}{c_{{3}}}^{2}{b_{{1}}}^{2}{t}^{2}-\\
 && 24\,\Lambda\,{c_{{1}}}^{2}{c_{{3}}}^{2}{b_{{2}}}^{2}{t}^{2}\beta+12\,\beta\,\Lambda\,{c_{{2}}}^{2}{c_{{3}}}^{2}{b_{{1}}}^{2}{t}^{2}+12\,\beta\,\Lambda\,{c_{{1}}}^{2}{c_{{2}}}^{2}{b_{{3}}}^{2}{t}^{2}-\\
 && \left.\Lambda\,{c_{{1}}}^{2}{c_{{3}}}^{2}{b_{{2}}}^{2}{t}^{2}\alpha\right)/\left(18\, G{c_{{1}}}^{2}{c_{{3}}}^{2}\alpha\,\beta\right)\\
%\end{eqnarray*}
%\begin{eqnarray*}
&&\frac{d^{4}}{dt^{4}}\hat{h}_{3}=  \left(-24\,\beta\,\Lambda\,{c_{{1}}}^{2}{c_{{2}}}^{2}y_{{9}}-\alpha\,\Lambda\,{c_{{1}}}^{2}{c_{{2}}}^{2}y_{{9}}-\alpha\,\Lambda\,{c_{{2}}}^{2}{c_{{3}}}^{2}y_{{1}}+12\,\Lambda\,{c_{{1}}}^{2}{c_{{3}}}^{2}y_{{5}}\beta-\Lambda\,{c_{{1}}}^{2}{c_{{3}}}^{2}y_{{5}}\alpha+\right.\\
 && 12\,\beta\,\Lambda\,{c_{{2}}}^{2}{c_{{3}}}^{2}y_{{1}}-2\,\alpha\,\Lambda\,{c_{{2}}}^{2}{c_{{3}}}^{2}b_{{1}}c_{{1}}t-2\,\alpha\,\Lambda\,{c_{{1}}}^{2}{c_{{2}}}^{2}b_{{3}}c_{{3}}t-48\,\beta\,\Lambda\,{c_{{1}}}^{2}{c_{{2}}}^{2}b_{{3}}c_{{3}}t+\\
 && 24\,\beta\,\Lambda\,{c_{{2}}}^{2}{c_{{3}}}^{2}b_{{1}}c_{{1}}t+24\,\Lambda\,{c_{{1}}}^{2}{c_{{3}}}^{2}b_{{2}}c_{{2}}t\beta-2\,\Lambda\,{c_{{1}}}^{2}{c_{{3}}}^{2}b_{{2}}c_{{2}}t\alpha+{c_{{1}}}^{2}{c_{{2}}}^{2}y_{{11}}\alpha+\\
&&  6\,\beta\,{c_{{1}}}^{2}{c_{{3}}}^{2}y_{{7}}+\alpha\,{c_{{1}}}^{2}{c_{{3}}}^{2}y_{{7}}+6\,{c_{{2}}}^{2}{c_{{3}}}^{2}y_{{3}}\beta-24\,{c_{{1}}}^{2}{c_{{2}}}^{2}{b_{{3}}}^{2}\beta+2\,{c_{{2}}}^{2}{c_{{3}}}^{2}{b_{{1}}}^{2}\alpha+\\
 && 12\,{c_{{2}}}^{2}{c_{{3}}}^{2}{b_{{1}}}^{2}\beta+{c_{{2}}}^{2}{c_{{3}}}^{2}y_{{3}}\alpha+2\,\alpha\,{c_{{1}}}^{2}{c_{{3}}}^{2}{b_{{2}}}^{2}+2\,{c_{{1}}}^{2}{c_{{2}}}^{2}{b_{{3}}}^{2}\alpha+12\,\beta\,{c_{{1}}}^{2}{c_{{3}}}^{2}{b_{{2}}}^{2}-\\
 && 12\,{c_{{1}}}^{2}{c_{{2}}}^{2}y_{{11}}\beta-\alpha\,\Lambda\,{c_{{1}}}^{2}{c_{{2}}}^{2}{b_{{3}}}^{2}{t}^{2}-\alpha\,\Lambda\,{c_{{2}}}^{2}{c_{{3}}}^{2}{b_{{1}}}^{2}{t}^{2}+12\,\Lambda\,{c_{{1}}}^{2}{c_{{3}}}^{2}{b_{{2}}}^{2}{t}^{2}\beta+\\
 && \left.12\,\beta\,\Lambda\,{c_{{2}}}^{2}{c_{{3}}}^{2}{b_{{1}}}^{2}{t}^{2}-24\,\beta\,\Lambda\,{c_{{1}}}^{2}{c_{{2}}}^{2}{b_{{3}}}^{2}{t}^{2}-\Lambda\,{c_{{1}}}^{2}{c_{{3}}}^{2}{b_{{2}}}^{2}{t}^{2}\alpha\right)/\left(18\, G{c_{{2}}}^{2}{c_{{1}}}^{2}\alpha\,\beta\right)
  \end{eqnarray*}

% Produces the bibliography via BibTeX.
\bibliographystyle{elsarticle-num}

\end{document}